\RequirePackage{fix-cm}
\documentclass[smallextended]{svjour3}       
\smartqed  

\usepackage{graphicx}

\usepackage{algorithm}

\usepackage{newfloat}
\usepackage{listings}

\usepackage{times}
\usepackage{latexsym}

\usepackage[T1]{fontenc}

\usepackage[utf8]{inputenc}

\usepackage{microtype}

\usepackage{inconsolata}

\usepackage{microtype}
\usepackage{graphicx}
\usepackage{subfigure}
\usepackage{booktabs} 

\usepackage{multirow}
\usepackage{graphicx}
\usepackage{textcomp}
\usepackage{xcolor}
\usepackage{subfigure}
\usepackage{colortbl}
\usepackage{pifont}
\usepackage{multirow}
\usepackage{amsmath,siunitx,booktabs}
\usepackage{framed}
\usepackage{float}
\usepackage{fancyhdr}
\usepackage{url}
\usepackage{pifont}
\usepackage[noend]{algpseudocode}
\usepackage{algorithm}
\usepackage{xspace}
\usepackage{paralist}
\usepackage{pifont}
\usepackage{tcolorbox}
\usepackage{todonotes}
\usepackage{enumitem}
\usepackage{listings}
\usepackage[switch]{lineno}

\usepackage{amssymb}
\usepackage{mathtools}
\newtheorem{myDef}{Definition}
\usepackage{bibentry}

\begin{document}

\title{GenCode: A Generic Data Augmentation Framework for Boosting Deep Learning-Based Code Understanding
}


\author{Zeming Dong$^{1}$,  Qiang Hu*$^{2}$, Xiaofei Xie$^{3}$, Maxime Cordy$^{1}$, Mike Papadakis$^{1}$, Yves Le Traon$^{1}$ and Jianjun Zhao$^{4}$ \\
	\normalsize $^{1}$University of Luxembourg, Luxembourg\\
        \normalsize $^{2}$ Tianjin University, China\\
	\normalsize $^{3}$Singapore Management University, Singapore\\
        \normalsize $^{4}$Kyushu University, Japan\\ 
}

\authorrunning{Dong and Hu et.al}

\date{Received: date / Accepted: date}

\maketitle

\begin{abstract}
Pre-trained code models lead the era of code intelligence, with multiple models designed with impressive performance. However, one important problem, data augmentation for code data that automatically helps developers prepare training data lacks study in this field. In this paper, we introduce a generic data augmentation framework, GenCode, to enhance the training of code understanding models. Simply speaking, GenCode follows a generation-and-selection paradigm to prepare useful training code data. Specifically, it employs code augmentation techniques to generate new code candidates first and then identifies important ones as the training data by influence scores. To evaluate the effectiveness of GenCode, we conduct experiments on four code understanding tasks~(e.g., code clone detection) and three pre-trained code models~(e.g., CodeT5) and two recent released code-specific Large Language Models (LLMs)~(e.g., Qwen2.5-Coder). Compared to the \emph{state-of-the-art (SOTA)} code augmentation method MixCode, GenCode produces pre-trained code models with 2.92\% higher accuracy and 4.90\% adversarial robustness on average. For code-specific LLMs, GenCode achieves an average improvement of 0.93\% in accuracy and 0.98\% in natural robustness.

\end{abstract}

\section{Introduction}

\emph{Deep learning~(DL)} for code has been one of the successful DL applications over the years. The methodology of applying deep learning to solve \emph{software engineering (SE)} tasks, such as software vulnerability detection~\cite{zhou2019devign}, is commonly known as \emph{code learning}. From the model level, different DL models specifically designed for code learning have been proposed, such as \emph{CodeBERT}~\cite{feng2020codebert}, \emph{CodeT5}~\cite{wang-etal-2021-codet5}, and \emph{Qwen2.5-Coder}~\cite{hui2024qwen2}. From the commercial perspective, multiple related products such as Copilot~\footnote{\url{https://github.com/features/copilot}} and Tabnine~\footnote{\url{https://www.tabnine.com}} have been widely used to help our daily development. As a result, in different downstream learning-based code understanding tasks, e.g., code clone detection~~\cite{lu2021codexglue} and program classification~\cite{puri2021codenet}, DL has already achieved significantly better results than traditional SE techniques. 

However, most of the existing code learning research focuses on designing new code representation techniques or code model architectures to further improve the effectiveness of DL for code, one important problem -- how to prepare high-quality training data gained limited attention even though recent research demonstrated it highly affects the performance of code models~\cite{sun2022importance}. The challenge behind this is preparing training data for code models is heavy work. Although today developers can easily collect raw code data from the open-source community such as GitHub~\footnote{\url{https://github.com}}, data labeling is still expensive, especially for code-related tasks. It requires huge human effort and professional domain knowledge.

To tackle this problem, data augmentation is a famous technique that increases the diversity and quality of training data based only on existing labeled data. In the code learning field techniques such as code refactoring methods and text-oriented data augmentation methods can be directly used for code augmentation. Unfortunately, existing studies~\cite{bielik2020adversarial,yu2022data} stated that code refactoring methods have limited advantages in the accuracy improvement of code models. Besides, the recent work~\cite{dong2025boosting} revealed that code models trained using existing data augmentation methods still suffer from the low robustness issue. Therefore, exploring more effective and reliable code augmentation methods is an emerging task.  

To bridge this gap, inspired by the great success of the typical two-stage  \emph{Automated Data Augmentation (AutoDA)} framework in the field of \emph{computer vision}~\cite{cubuk2020randaugment,yang2023survey}, we propose GenCode, the first generation-and-selection-based data augmentation framework for code understanding. Specifically, at each training epoch, GenCode first generates~\emph{training candidates} by using different code augmentation methods. Here, we consider different types of methods such as semantic-preserving code refactoring methods and syntax-breaking text augmentation methods. After that, GenCode selects valuable training data from the candidates with the guidance of influence scores. The influence score used in GenCode is simply defined as the \emph{loss value} of code data produced by the code model.  Finally, the selected samples are used for the model training. We conducted comprehensive experiments to study how GenCode can improve the accuracy and robustness of code models. Based on that, we answer the following three research questions:

\smallskip
\noindent\textbf{RQ1:How effective is GenCode in producing accurate code models?} The results demonstrate that, on average, GenCode can enhance the accuracy of pre-trained code models by up to 3.63\% and 2.92\% compared to the training without using data augmentation and the SOTA method MixCode, respectively. For code-specific LLMs, Gencode achieves average improvements of 0.93\% for StarCoder2 and 0.84\% for Qwen2.5-Coder. Notably, our statistical tests show GenCode consistently outperforms baseline methods in terms of accuracy improvement.

\smallskip
\noindent\textbf{RQ2: How effective is GenCode in producing robust code models?} The results demonstrate that GenCode produces more robust pre-trained code models compared to existing data augmentation methods. Specifically, GenCode achieves an average robustness improvement of 8.42\% over models trained without data augmentation and 4.90\% over models trained with MixCode. However, for code-specific LLMs, the robustness gains are modest, with improvements of less than 1\%.

\smallskip
\noindent\textbf{RQ3: How does the influence score affect the effectiveness of GenCode? }
The results suggest that using the influence score based on maximum loss values is the most effective strategy among the considered variants, including minimum loss value selection and random selection. Besides, even GenCode with random selection outperforms existing code augmentation methods.

\smallskip
\noindent\textbf{RQ4: How does the data selection method affect the effectiveness of GenCode?} Experimental results indicate that GenCode achieves superior performance in terms of accuracy improvement. However, the gradient-based method outperforms GenCode and other data selection methods in enhancing model robustness.


To summarize, the main contributions of this paper are:

\begin{itemize}

\item We propose the first generation-and-selection-based framework, GenCode~\footnote{\url{https://github.com/zemingd/GenCode}\label{site_GenCode}}, which automatically enriches the training data for boosting the code understanding models. 
\item We empirically demonstrate that GenCode can produce more accurate and robust code understanding models compared to existing data augmentation methods.

\item We conduct an ablation study to explore how different influence scores and data selection methods affect the effectiveness of GenCode.

\end{itemize}

\section{Background}
\label{sec:background}

In this section, we introduce the necessary knowledge that is needed to understand our work, code representation, data augmentation for code understanding, and robustness testing of code model.
\subsection{Code Representation}
Code representation is a crucial technique that converts source code into a format interpretable by \emph{Deep Neural Networks (DNNs)}, facilitating effective learning of source code features. Since source code can be seen as sequences of tokens or structural graph~\cite{allamanis2018survey}, researchers employ the techniques in \emph{Natural Language Processing (NLP)} and \emph{Graph Neural Networks (GNNs)} to learn information from code and solve downstream tasks, such as automated program classification~\cite{puri2021codenet} and code clone detection~\cite{wang2020detecting}. 
Existing methods~\cite{allamanis2018survey} for code representation can be roughly divided into two groups, sequence-based representation and graph-based representation.
\textit{Sequence-based representation} follows the same procedure of text data processing in NLP to convert raw code data into a sequence of code tokens~\cite{alon2018codeseq}. 
\textit{Graph-based representation} learns the structural information of code. Here, the structural information can be \emph{abstract syntax tree~(AST)}, data flow information, and other types of flow information~\cite{dinella2020hoppity,yasunaga2020graph}. 

In this work, we focus on code data preprocessing and study the powerful pre-trained code models and code-specific LLMs that achieved impressive results on code understanding tasks. 
 
\subsection{Data Augmentation for Code Understanding} 

Data augmentation is a well-studied technique in the machine learning field that increases the size and diversity of training data by only modifying the existing labeled data, thus, without further human effort. For instance, in the NLP field, researchers tried many ways to augment the text data, e.g., replacing a token with a synonym~\cite{feng2021survey}.

Recently, data augmentation has gained attention from software engineering researchers. Various data augmentation methods designed for source code have been proposed~\cite{gao2020fuzz,yefet2020adversarial,allamanis2021self,pour2021search,bui2021self,wang2022bridging,park-etal-2023-contrastive,yang2023exploitgen,dong2024effectiveness_JSS}, with the most general framework involving the use of code refactoring, e.g., \emph{local variable renaming} and \emph{dead store}, to create more code data without breaking the semantics of the original codes. Concretely, code refactoring rewrites the syntactic structure of source code while keeping the semantic information~\cite{lacerda2020code}. 

Existing works~\cite{bielik2020adversarial,yu2022data,de2023detecting,dong2025boosting,dong2024effectiveness} conducted empirical studies and found code refactoring methods have limited advantages in boosting the training of code models. For instance, a recent empirical study~\cite{dong2024effectiveness}  on code understanding revealed that the utilization of code refactoring methods only achieved success rates of 66.67\% and 41.67\% in improving accuracy and robustness, respectively, compared to code model training without data augmentation. Besides, the SOTA code augmentation method, MixCode, which linearly interpolates pairs of source codes, has been proven more effective than the current general framework code refactoring. However, as a variant of Mixup specifically designed for code data, MixCode is limited to applications in code classification tasks, as it necessitates the label with one-hot encoding.

However, existing studies~\cite{bielik2020adversarial,yu2022data} have stated that these code refactoring methods have limited advantages in boosting the training of code models. To solve this problem, we propose a simple yet effective code augmentation framework that is not only unaffected by the dataset and model constraints but also applicable to both classification and non-classification tasks.

\subsection{Robustness Testing of Code Model}
Robustness is an important property of DL models which indicates how the DL model can handle data with different distribution of the training data.
Normally, there are two types of robustness of DL models, natural robustness and adversarial robustness. Natural robustness indicates how models perform when facing the natural distribution shift, for example, data in different environments and newly collected data. Adversarial robustness refers to the ability of models to defend the adversarial attacks. Robustness testing of code models has been addressed recently. Researchers proposed distribution shifted code~(e.g., codes have the same functionality but are written by different programmers) data to study the natural robustness of code models. Besides, multiple adversarial attack methods have been proposed to test the adversarial robustness of code models, for example, the recent work~\cite{yang2022natural} proposed to change the variable name of code to force the model to make wrong predictions. 

In this work, we employ adversarial attack methods~\cite{dong2025boosting} proposed for code to evaluate the adversarial robustness of our trained models. Following the existing work~\cite{dong-mixcode}, we construct the natural robustness datasets to evaluate the performance of all code models.



\section{Definition} 

We first define the problem of data augmentation for code models. 

\begin{myDef}[Program Code] \textit{A program code \footnote{For simplicity, we use \textit{code} in the paper.}  is defined as a 5-tuple $\mathcal{P} = \left(content, se, sy, Ip, Op\right)$: $content$ contains all the text in the code $\mathcal{P}$; $se$ and $sy$ are the semantic and syntax of the code $\mathcal{P}$; and $Ip$ and $Op$ are the input and output of the code $\mathcal{P}$.}
\end{myDef} 

\begin{myDef}[Code Model] \textit{A code model is defined as a 3-tuple $M = \left(parameter, \mathcal{P}, Y\right)$, where $parameter$ contains all the weights in code model $M$, $\mathcal{P}$ is the input code, and $Y$ is the label of $\mathcal{P}$ given a specific downstream code task.}
\end{myDef}

Code models can be any machine learning model such as a simple regression model or a pre-trained model, e.g., CodeT5~\cite{wang-etal-2021-codet5}.

\begin{myDef}[Code Augmentation]
\label{def:codeaug}
\textit{Given a code $P$, a code augmentation method $\mathcal{T}: \mathcal{X} \rightarrow \mathcal{X'}$, code augmentation is to transform $\mathcal{P}$ to $\mathcal{P'}$ that $\mathcal{P'}.content = \mathcal{T}\left(\mathcal{P}.content\right) \& \mathcal{P'}.content \neq \mathcal{P}.content$.} 
\end{myDef}

Here, code augmentation methods can be any method that can change the content of program statements such as text-oriented data augmentation methods (e.g., synonym replacement and code refactoring methods (e.g., dead if statement adding).

\begin{myDef}[Semantic-Preserving~(Breaking)]
\label{def:sem}
\textit{Given a code model $M$, a code $\mathcal{P}$ and its augmented version $\mathcal{P'}$, semantic-preserving (breaking) code augmentation indicates $\mathcal{P}.se = \mathcal{P'}.se$ \& $M(\mathcal{P}) = M(\mathcal{P'})$ ($\mathcal{P}.se \neq \mathcal{P'}.se$ \&  $M(\mathcal{P}) \neq M(\mathcal{P'})$).}  
\end{myDef}

\begin{myDef}[Syntax-Preserving~(Breaking)]
\label{def:syn}
\textit{Given a code $\mathcal{P}$ and its augmented version $\mathcal{P'}$, syntax-preserving (breaking) code augmentation indicates $\mathcal{P}.sy = \mathcal{P'}.sy$ ($\mathcal{P}.sy \neq \mathcal{P'}.sy$).}
\end{myDef}

\section{Assumption and Preliminary Study}
\label{sec:intuition}

GenCode utilizes the influence score to select the most valuable data for the code model training, where the influence score is defined as the loss value. Therefore, this approach is grounded on the assumption that the data that code models have less confidence in is more useful for model training. To check whether this assumption is valid or not, we conducted a preliminary study to explore if there is a correlation between the accuracy of the trained code model and the confidence of the model on the training data. Specifically, the study is conducted on the \emph{Google Code Jam (GCJ)}~\cite{yang2022natural} dataset and CodeBERT pre-trained code model. The detailed steps are as follows:

\textbf{Step 1.} We initialize the code model and use code augmentation methods to generate additional data as the training candidate set. The code augmentation methods will be introduced in a later section.

\textbf{Step 2.} We randomly select a number of data from the training candidate set. Here the selected number is the same as the size of the original training data. We repeat this selection 100 times and collect 100 groups of training data. After that, we compute and record the average loss values for each group of training data.

\textbf{Step 3.} We train the code model using each group of training data respectively for a few epochs and thus save 100 trained code models. We then test each code model using the original test data and save 100 accuracies.

\textbf{Step 4.} We use the \emph{Pearson Correlation Coefficient (PCC)}~\cite{schober2018correlation} metric to check the correlation between the loss values and the accuracies. 

\begin{figure*}[h]
\centering
\subfigure[Initialized]{
\begin{minipage}[t]{0.5\linewidth}
\includegraphics[width=1.0\linewidth]{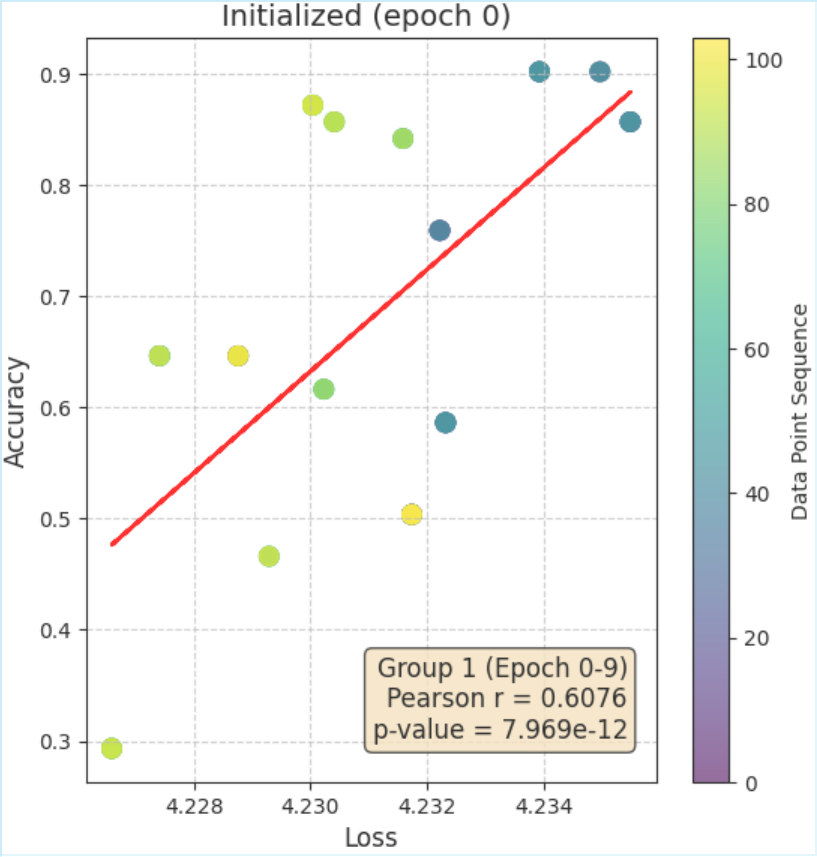}
\end{minipage}%
}%
\subfigure[Middle]{
\begin{minipage}[t]{0.5\linewidth}
\includegraphics[width=1.0\linewidth]{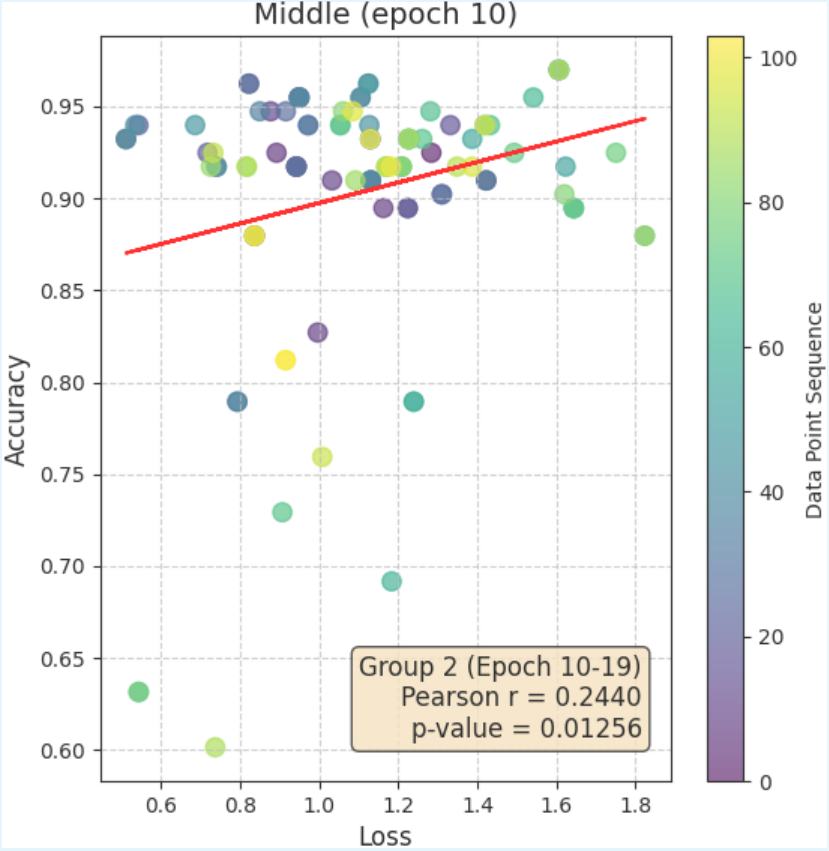}
\end{minipage}%
}%

\subfigure[Well-trained]{
\begin{minipage}[t]{0.5\linewidth}
\centering
\includegraphics[width=1.0\linewidth]{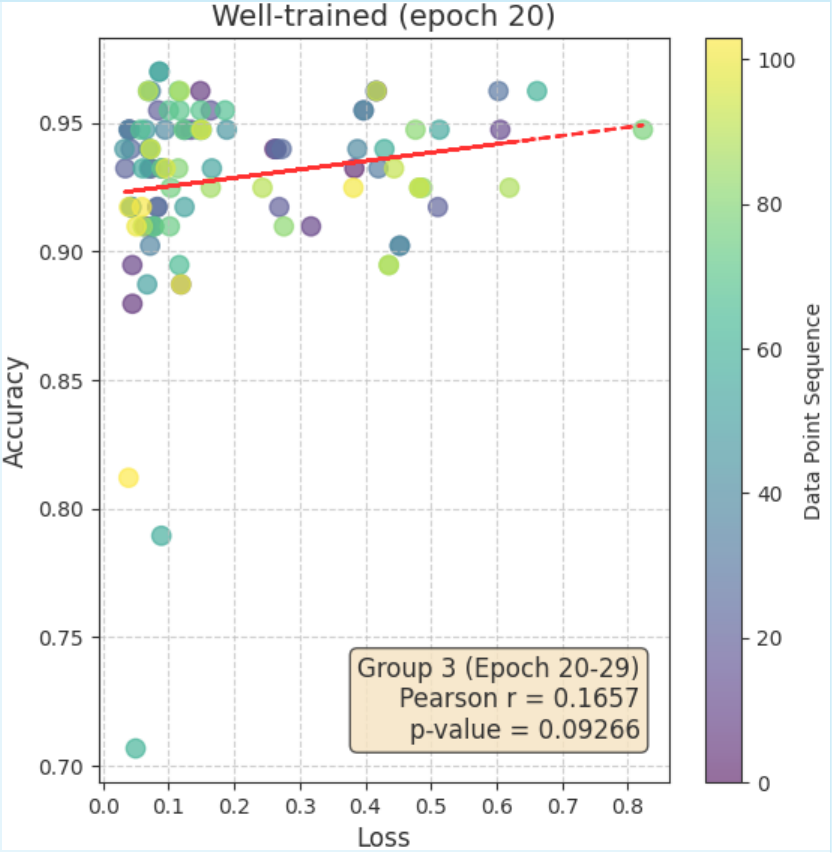}
\end{minipage}%
}%
\caption{Correlation between loss values and code model accuracy.}
\label{fig:correlation}
\vspace{-3mm}
\end{figure*}

We consider three training stages, which are the Early stage (Initialized, Epochs 0–9), the Mid stage (Middle, Epochs 10–19), and the Late stage (Well-trained, Epochs 20-29). Fig.~\ref{fig:correlation} visualizes the results of our preliminary study. Here, we considered three types of code models, the initialized model, the model that has been trained for a few epochs but not converged, and the well-trained code model, to check if our assumption can stand in different scenarios.

 The results indicate that there is a positive correlation between the loss values of data and the accuracy of code models trained by using these data. From the initialized model to the well-trained model, this correlation becomes weaker but still exists. We conjecture that when the model is in the early stage, most of the data that have high loss values with more informative and lead to better model generalization. However, when training proceeds, the data that still have high loss values are the model itself cannot learn. As training progresses, the correlation between loss values and accuracy gradually weakens, which may be attributed to overfitting or the model's limited capacity to learn from inherently difficult samples. This observation supports the hypothesis that the strength of this correlation is indicative of the model’s generalization ability. In particular, the decline in correlation from early to later training stages highlights the shifting informativeness of high-loss samples. Leveraging this insight, we design a data augmentation framework that prioritizes informative samples, especially in the early stages, to enhance the training effectiveness of code models.
 
 \begin{tcolorbox}
\textbf{Finding}: There is a positive correlation between the loss values of code data and the accuracy of code models trained by using these data. When the code model is randomly initialized, this correlation is significant~(with a \emph{P-value} less than 0.01).
\end{tcolorbox}

\section{Methodology}
\label{sec:methodology}

\subsection{Overview}

The goal of GenCode is to produce code models with better performance on test data by training models on carefully generated data. Given a training set of program $\mathcal{P}_{train}$, its transformed version $\mathcal{P'}_{train}$, and a set of test programs $\mathcal{P}_{test}$, after the training process $  \mathsf{train} \left(\cdot\right)$ , $\varrho(M'(\mathcal{P}_{test})) > \varrho(M(\mathcal{P}_{test}))$, where $M.parameter = \mathsf{train}(M, \mathcal{P}_{train})$ \& $M'.parameter = \mathsf{train}(M', \mathcal{P'}_{train})$, and $\varrho \left(\cdot\right)$  is the performance measurement. 

\begin{figure*}[h]
    \centering
    \includegraphics[scale=0.4]{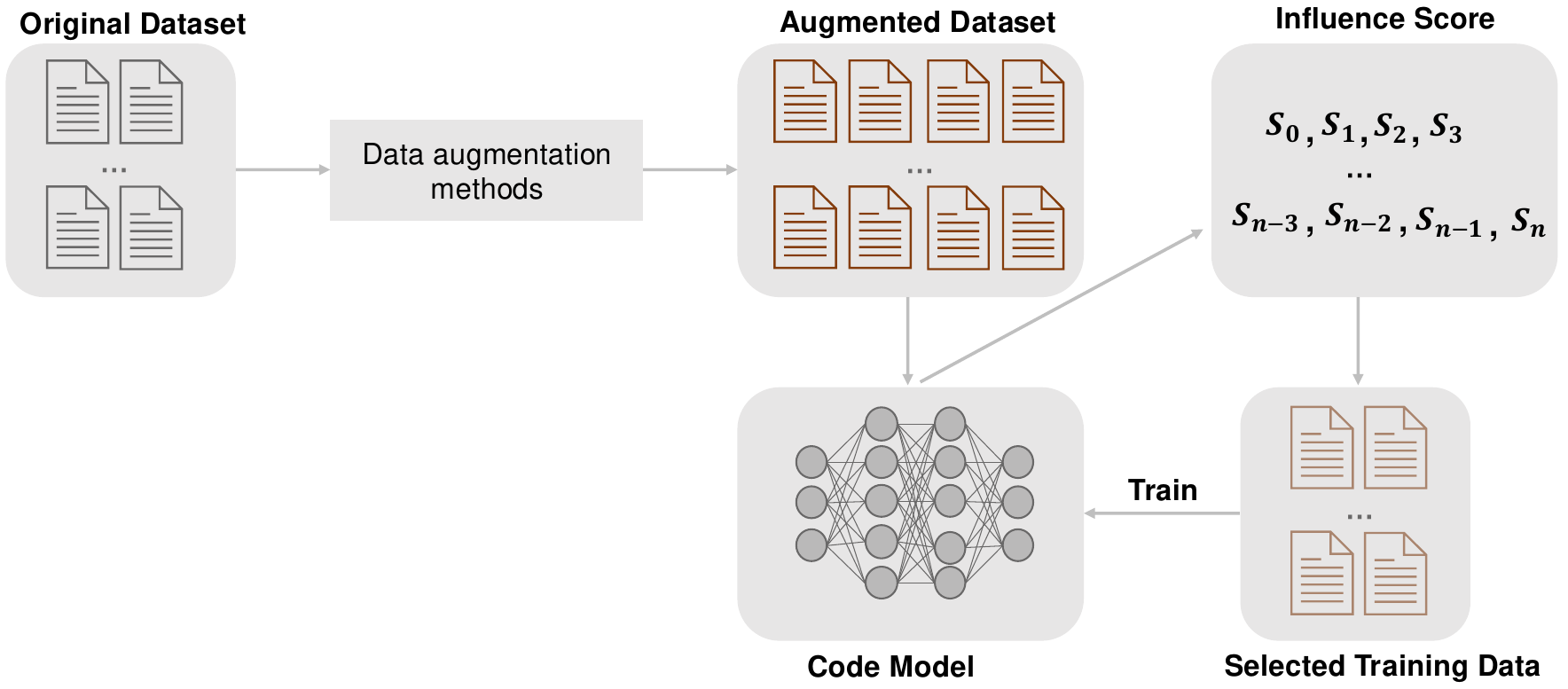}
    \caption{Workflow of GenCode in one training epoch.}
    \label{fig:workflow}
    \vspace{-3mm}
\end{figure*}

GenCode follows the generation-and-selection workflow and utilizes different types of code augmentation methods to generate diverse code data first, then selects code data that are most useful for training as the training data. Algorithm~\ref{alg:ida} shows how GenCode works. In each training epoch, GenCode automatically generates more code data~(Line~\ref{line: startgen} to Line~\ref{line: endgen}, Definition~\ref{def:codeaug}) and then selects the most important ones~(Line~\ref{line: lossstart} to Line~\ref{line: lossend}) to train the code model~(Line~\ref{line: train}). Here, the data importance is measured by the loss values produced by the code model which is the most common indicator to check how the model performs on the data. 
\begin{algorithm}[H]
\caption{GenCode}
\label{alg:ida}
\begin{algorithmic}[1]
\Require $M$: initialized DNN model
\Require $\mathcal{P}, Y$: original code data and labels
\Require $\mathcal{T}=\left\{t\right\}:$ a set of code augmentation methods
\Ensure $M$: trained model
\For{$run\gets \{0, \ldots, \#epochs\}$}
\State $\mathcal{P'}, Y'\ \gets \phi, \phi$ \Comment{initialization} 

\For{$t\in \mathcal{T}$} \label{line: startgen}
\State $\mathcal{P'}\gets \mathcal{P'}\cup t\left(\mathcal{P}\right)$
\State $Y'\gets Y'\cup Y$  \label{line: endgen}
\Comment{code augmentation}

\EndFor
\State $I\gets \mathsf{Importance}\left(M, \mathcal{P'}\right)$ \label{line: lossstart}
\Comment{computing importance}
\State $Index\gets \mathsf{argSort}\left(I\right)\left[:||\mathcal{P}||\right]$
\Comment{sort by importance}
\State $X_{train}\gets \mathcal{P'}\left[Index\right]$ 
\State $Y_{train}\gets Y'\left[Index\right]$ \label{line: lossend}
\State $M\gets \mathsf{Train}\left(M, X_{train}, Y_{train}\right)$ \label{line: train}
\EndFor
\State \Return $M$
\end{algorithmic}
\end{algorithm}
More specifically, Fig.~\ref{fig:workflow} depicts the workflow of GenCode in one training epoch. First, a series of code augmentation methods are applied to the original dataset to generate an augmented dataset (\emph{search space}). Then, samples are selected based on their influence score, i.e., the loss during model training (\emph{importance-guided data selection}), ensuring the selected dataset keeps the same size as the original dataset for a fair comparison.

Overall, GenCode contains two main steps, utilizing code augmentation methods to compose the search space and importance-guided data selection (will be introduced in Section~\ref{subsec:search space} and Section~\ref{subsec:Importance-Guided Data Selection}, respectively). 

\subsection{Search Space}
\label{subsec:search space}
Even though the code has strict syntax constraints,  recent work~\cite{dong2025boosting} found that these data augmentation methods that slightly break the syntax of the source code e.g., \emph{Random Swap}, can still be useful for code model training.  To enhance the diversity of source code within the feature space, GenCode considers two types of code augmentation methods, semantic/syntax-preserving methods, and syntax-breaking methods~(see Definition~\ref{def:sem} and Definition~\ref{def:syn}). For instance, synonym replacement (SR) may turn an identifier ''\verb|count|'' into a natural-language word, such as ''\verb|amount|'', which is not declared in the symbol table and thus results in a compilation error. In contrast, refactoring-based renaming updates all references consistently and guarantees syntactic correctness.

\begin{figure}[h]
    \centering
    \includegraphics[scale=0.4]{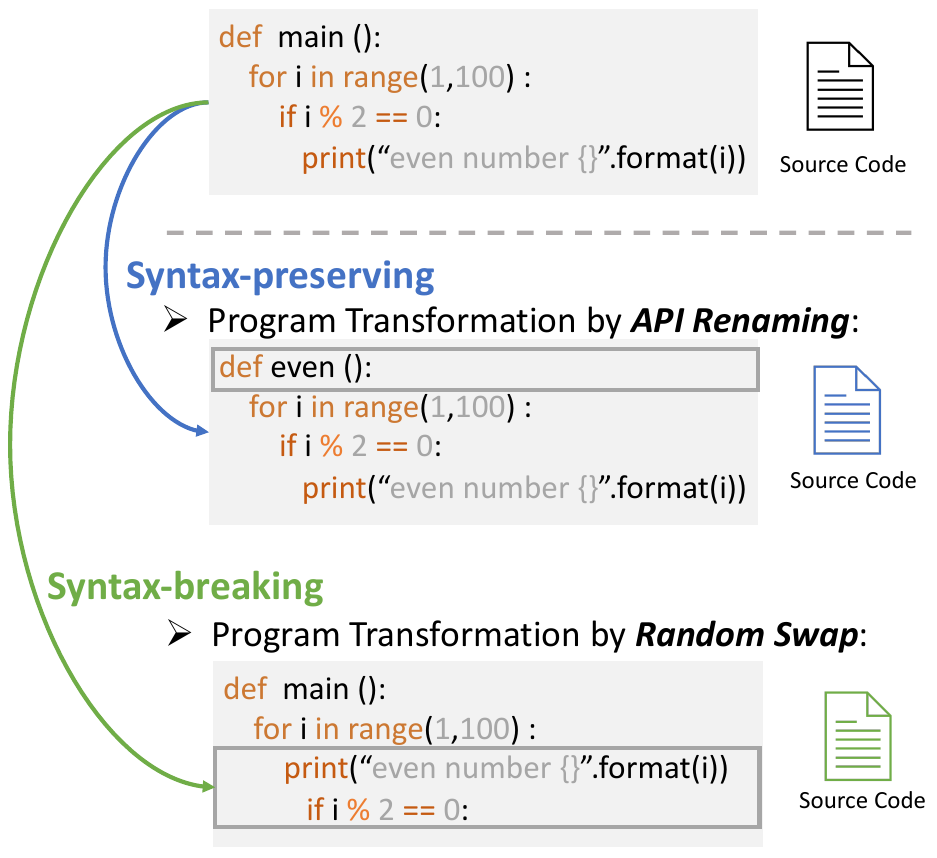}
    \caption{An example of atomic data augmentation operators.}
    \label{fig_example}
\end{figure}

GenCode adapts code refactoring methods (i.e., API renaming and Duplication) as semantic/syntax-preserving methods. Code refactoring methods follow the constraints of programming languages (PL) to edit the code while keeping the main functionality of the code. On the other hand, text-oriented data augmentation methods~\cite{dong2025boosting} are used as syntax-breaking methods in GenCode, which are widely used in the NLP field. For example, text-oriented data augmentation methods can randomly delete some tokens in the code. However, even though the syntax has been broken, the code model can still successfully predict those codes since it reads the code as a sequence. In total, as shown in Table~\ref{table_da4code}, GenCode currently supports 18 code refactoring methods and 5  text-oriented data augmentation methods. Besides, Fig.~\ref{fig_example} provides an example of the training candidate that is to generate the search space. In the syntax-preserving, we present a code refactoring method named \emph{API Renaming}, which involves modifying the function name from ''\verb|main()|''  to ''\verb|even()|''. In the syntax-breaking, we demonstrate another instance of~\emph{Random Swap}, where two different statements within a program are randomly selected and their positions swapped. As depicted in this example, the third and fourth lines have been interchanged. 


\begin{table}[!tb]
\caption{Atomic data augmentation operators adopted in GenCode. }
\label{table_da4code}
\resizebox{\columnwidth}{!}{
\begin{tabular}{ll|c}
\cline{1-3}
\multicolumn{2}{c}{Syntax-preserving} & \multicolumn{1}{c}{Syntax-breaking}  \\ \cline{1-3}
\multicolumn{1}{l}{\begin{tabular}[l]{@{}l@{}}API renaming\end{tabular}} & \multicolumn{1}{l|}{\begin{tabular}[l]{@{}l@{}}Field enhancement\end{tabular}} & \multirow{9}{*}{\begin{tabular}[c]{@{}l@{}} \multicolumn{1}{l}{Synonym replacement (SR)}   \\  \multicolumn{1}{l} {Random insertion (RI)}   \\  \multicolumn{1}{l}{Random swap (RS)} \\  \multicolumn{1}{l}{Random deletion (RD)} \\  \multicolumn{1}{l}{Back-translation (BT)} \end{tabular}}  \\ 
\multicolumn{1}{l}{\begin{tabular}[c]{@{}l@{}}Arguments adding\end{tabular}} & \multicolumn{1}{l|}{\begin{tabular}[c]{@{}l@{}}For loop enhancement\end{tabular}}       &      \\ 
\multicolumn{1}{l}{\begin{tabular}[c]{@{}l@{}}Arguments renaming\end{tabular}} & \multicolumn{1}{l|}{\begin{tabular}[c]{@{}l@{}}If enhancement\end{tabular}}         &        \\ 
\multicolumn{1}{l}{\begin{tabular}[c]{@{}l@{}}Dead for adding\end{tabular}} & \multicolumn{1}{l|}{\begin{tabular}[c]{@{}l@{}}Local variable adding\end{tabular}}       &        \\ 
\multicolumn{1}{l}{\begin{tabular}[c]{@{}l@{}}Dead if adding\end{tabular}} & \multicolumn{1}{l|}{\begin{tabular}[c]{@{}l@{}}Local variable renaming\end{tabular}}         &      \\ 
\multicolumn{1}{l}{\begin{tabular}[c]{@{}l@{}}Dead if-else adding\end{tabular}} & \multicolumn{1}{l|}{\begin{tabular}[c]{@{}l@{}}Method name renaming \end{tabular}}        &      \\ 
\multicolumn{1}{l}{\begin{tabular}[c]{@{}l@{}}Dead switch adding\end{tabular}} & \multicolumn{1}{l|}{\begin{tabular}[c]{@{}l@{}}Plus zero\end{tabular}}       &      \\ 
\multicolumn{1}{l}{\begin{tabular}[c]{@{}l@{}}Dead while adding\end{tabular}} & \multicolumn{1}{l|}{\begin{tabular}[c]{@{}l@{}}Print adding\end{tabular}}       &      \\ 
\multicolumn{1}{l}{Duplication} & \multicolumn{1}{l|}{Return optimal}   &   \\ \cline{1-3}
\end{tabular}
} 
\vspace{-3mm}
\end{table}

\subsection{Importance-Guided Data Selection}
\label{subsec:Importance-Guided Data Selection}

After constructing the training candidate, we then choose the most valuable code data that can benefit the code model training. Based on our \emph{Finding} drawn from the preliminary study, GenCode chooses the simplest way to measure the importance of code data to the model -- the \emph{loss} value. That means, GenCode takes all the transformed code as input to the code model and records all the loss values first, then ranks all the code in descending order based on the loss values and selects the first $K$ code as the training data. Following the common setting in the data augmentation problem~\cite{dong-mixcode}, $K$ is set as the number of original training data. GenCode chooses loss value as the indicator since it can be generalized to any code model and task.

\section{Evaluation Setup}
\label{sec:setup}
We evaluate both the accuracy and robustness of trained code models. Besides, we check how does influence score affects the effectiveness of GenCode.

\subsection{Study Design}
We conduct experiments to evaluate the usefulness of GenCode and answer the following research questions.

\textbf{RQ1:} \textit{How effective is GenCode in producing accurate code models?} Accuracy is the fundamental metric that is used to evaluate the performance of a trained model. Therefore, we first study the basic DNN property, the test accuracy, of each selected pre-trained code model. Specifically, we compare GenCode with models trained without using data augmentation and models trained by baseline data augmentation methods (will be introduced in Section~\ref{sec:baselines_GenCode}), and we additionally provide the evaluation results of text-oriented data augmentation methods (as mentioned in Section~\ref{subsec:search space}) to demonstrate the usefulness of GenCode. Additionally, to ensure a fair comparison in terms of training cost, we apply each data augmentation method to every sample in the original training dataset, resulting in an augmented dataset of the same size. These augmented datasets are fixed prior to training and remain unchanged across all training epochs.

\textbf{RQ2:} \textit{How effective is GenCode in producing robust code models?} Robustness is another important property of DNN models. In this study, we mainly focus on adversarial robustness and natural robustness. Specifically,  for the adversarial robustness, we utilize two widely used adversarial attack methods including MHM~\cite{zhang2020generating} and ALERT~\cite{yang2022natural}, to assess the performance of code models trained by the data augmentation method in robustness improvement. For natural robustness, we randomly apply a single code refactoring method~\cite{dong-mixcode} to each sample in order to construct a transformed code dataset that preserves the original semantic labels, thereby enabling a comprehensive evaluation of the generalization capability under natural variations.

\textbf{RQ3:} \textit{How does the influence score affect the effectiveness of GenCode?} The importance of score-based data selection is the key component is GenCode. Therefore, in this RQ, we change the selection strategies, i.e., selecting data with minimum confidence and randomly selecting data, to study the contribution of our proposed selection strategy. To achieve this, we repeat the experiments outlined in RQ1 and RQ2. Subsequently, we investigate how the selection of different scores influences the performance of pre-trained models. Note that, for a fair comparison, we evaluate only the adversarial robustness, as the use of \emph{random} may artificially inflate performance on the transformed dataset~\cite{dong-mixcode}.

\textbf{RQ4:} \textit{How does the data selection method affect the effectiveness of GenCode?} Data selection strategies play a critical role in determining the performance of trained code models~\cite{hu2024active}. In this RQ, we select representative methods from the literature to serve as baselines for comparison with GenCode. Specifically, we include \emph{LIME}~\cite{ribeiro2016should}, which provides interpretable, local approximations of complex model predictions and can be leveraged during the model training phase to inform data selection by identifying influential samples. As one of the classical data selection methods, \emph{Influence Function}~\cite{koh2017understanding} estimates the effect of individual training samples by approximating how infinitesimal perturbations to the training data impact the learned parameters, and can be effectively utilized during the phase of model training to guide data selection. Additionally, we incorporate \emph{BADGE}~\cite{ash2019deep}, which selects data by applying K-means clustering on gradient embeddings, and has been shown to be a stable and effective active learning method in recent empirical studies\cite{hu2024active}. We evaluate the performance of selected data selection methods in terms of both accuracy and natural robustness. Adversarial robustness is not considered in this study, as existing adversarial attacks in RQ2 are not well-suited for LLMs.

\subsection{Datasets and Models}
\label{sec:dataset_models}

We evaluate GenCode on four code understanding tasks as shown in Table~\ref{tab:data_model}, bug detection~\cite{hu2019re}, authorship attribution~\cite{yang2022natural}, code clone detection~\cite{yang2022natural} 
(Note that, MixCode is not suitable for this task since its input is code pairs.), and problem classification~\cite{puri2021codenet}. We conduct all experiments on a server equipped with 4 NVIDIA RTX A6000 GPUs. For each code task, we fine-tune three types of models based on pre-trained code models, CodeBERT, GraphCodeBERT, and CodeT5. For the model training, we follow existing works~\cite{hu2019re,wang-etal-2021-codet5,puri2021codenet,yang2022natural,dong-mixcode}, to set all configurations. Specifically, we set the training epoch as 50 for all tasks and models. Regarding the optimizer, we utilize \emph{Adam}~\cite{kingma2014adam} with a learning rate of $10^{-3}$ for all the models. 
Besides, we use early stopping with a patience of 20 to mitigate overfitting. To reduce the influence of randomness, we train each model five times and report the average results along with the standard deviation. For MixCode, we follow the recommendation of the original paper and set the Mixup ratio as $\alpha = 0.1$. For the setting of adversarial attacks, we directly use the default configurations provided by the original papers~~\cite{yang2022natural}. Considering the code-specific Large Language Models (LLMs), we select \emph{StarCoder2}~\cite{lozhkov2024starcoder}, a representative model from the \emph{BigCode}~\footnote{\url{https://www.bigcode-project.org}} project, which is a collaborative initiative dedicated to the responsible development of LLMs for Code, alongside the recently released \emph{Qwen2.5-Coder}~\cite{hui2024qwen2}.

\begin{table}[h]
\caption{Details of tasks, datasets, and DNN models. \textbf{\#Training} and \textbf{\#Test} are the number of data for training and testing, respectively.}
\label{tab:data_model}
\resizebox{.98\columnwidth}{!}{
\begin{tabular}{lllcccc}
\toprule
\textbf{Dataset} & \textbf{Language} & \textbf{Task} & \textbf{\#Training} & \textbf{\#Test} & \textbf{Model} \\ \hline
\multirow{3}{*}{Google Code Jam (GCJ)} & \multirow{3}{*}{Python} & \multirow{3}{*}{Authorship attribution} & \multirow{3}{*}{528} & \multirow{3}{*}{132} & \multirow{6}{*}{\begin{tabular}[c]{@{}c@{}} CodeBERT\\ GraphCodeBERT\\ CodeT5 \\ StarCoder2-7B\\ Qwen2.5-Coder-7B \end{tabular}} \\
&  &  &  &  &   \\ 
&  &  &  &  &   \\ \cline{1-5}
\multirow{3}{*}{Refactory} & \multirow{3}{*}{Python} & \multirow{3}{*}{Bug detection} & \multirow{3}{*}{3,380} & \multirow{3}{*}{423} &  \\
 &  &  &  &  &      \\ 
 &  &  &  &  &   \\  \hline 
\multirow{3}{*}{Java250} & \multirow{3}{*}{Java} & \multirow{3}{*}{Problem classification} & \multirow{3}{*}{48,000} & \multirow{3}{*}{15,000} & \multirow{6}{*} {\begin{tabular}[c]{@{}c@{}}  CodeBERT\\ GraphCodeBERT \\ CodeT5 \\ StarCoder2-7B \\ Qwen2.5-Coder-7B   \end{tabular}} \\
 &  &  &  &  &    \\
  &  &  &  &  &   \\ \cline{1-5}
\multirow{3}{*}{BigCloneBench} & \multirow{3}{*}{Java} & \multirow{3}{*}{Clone detection} & \multirow{3}{*}{90,102} & \multirow{3}{*}{4,000} &  \\
 &  &  &  &  &   \\
  &  &  &  &  &   \\
 \bottomrule
\end{tabular}
}
\vspace{-3mm}
\end{table}

\subsection{Configuration}

We conduct all experiments on a computation server equipped with 4× NVIDIA A100 (40GB) GPUs, two Intel Xeon Gold 6330 CPUs, and 512GB RAM. To ensure fairness and consistency across benchmarks, we adopt a unified training configuration following established practices in prior work~\cite{wang-etal-2021-codet5,dong-mixcode,hu2019re,yang2022natural,puri2021codenet}. Specifically, we train CodeBERT, GraphCodeBERT, and CodeT5 for 50 epochs, and train StarCoder2-7B and Qwen2.5-Coder-7B for 20 epochs, using the Adam optimizer (learning rate = 1e-3, weight decay = 0.01). The batch size is fixed to 32, and the maximum sequence length is set to 512 tokens for all encoder-based architectures. To mitigate overfitting and ensure stable convergence, we apply early stopping with a patience of 20 epochs. Each experiment is executed five independent times, and we report the mean and standard deviation to account for stochastic variability. For MixCode, we follow the settings recommended in the original paper and adopt a mixup coefficient of $\alpha = 0.1$. For adversarial training baselines, we directly use the default hyperparameters released by their authors to avoid unintended deviations from the original configurations.

To support reproducibility, we provide the complete training pipeline~\footref{site_GenCode}, including data preprocessing scripts, model configuration files (e.g., batch size, sequence length, and optimizer parameters), and execution instructions. This ensures that all experimental results presented in this work can be reproduced under the same computational environment and hyperparameter settings.

\subsection{Evaluation metrics and Baselines}
\label{sec:baselines_GenCode}
In the evaluation, we assess the performance of trained models from two perspectives, namely, \emph{clean accuracy} and \emph{model robustness}. For the robustness evaluation, we check the \emph{attack success rate (ASR)} metric of two well-established adversarial attack methods, namely, \emph{ALERT}~\cite{yang2022natural}, which uses variable renaming while considering natural semantics, and \emph{MHM}~\cite{zhang2020generating}, which also randomly replaces variable names but does not account for natural semantics. Metric \emph{ASR} evaluates the performance of models, including CodeBERT, GraphCodeBERT, and CodeT5~\cite{dong2025boosting}. In addition, we evaluate the natural robustness of all trained models by applying 18 different code refactoring methods to generate diverse program variants. Specifically, following the existing work~\cite{dong2025boosting,dong-mixcode}, we construct transformed versions of each code sample in the dataset by randomly applying one of the refactoring methods.


We consider \emph{model training without data augmentation (No Aug)}, \emph{Refactor}~\cite{yu2022data}, a family of text-oriented data augmentation methods~\cite{dong2025boosting}, and MixCode as baseline methods. \emph{Refactor}, which keeps code syntax rules, is a widely used data augmentation framework specifically designed for code data. In contrast, text-oriented data augmentation methods, which slightly break syntax rules, have recently proven more effective than Refactor. Additionally, MixCode is a more recent and high-performing code augmentation method that significantly enhances code models. 

\section{Results Analysis}
\label{sec:results}
\subsection{RQ1: How effective is GenCode in producing accurate code models?}
\label{sec:results_acc}

\textbf{Overall accuracy.}  Column~\emph{Accuracy} in Table~\ref{tab:ACC_pre-trained} summarizes the final test accuracy of code models trained by using different code augmentation methods including GenCode. Surprisingly, from the results, we can see that most of the existing code augmentation methods except \textit{RS} and MixCode negatively affect the performance of code models. This finding is similar to the conclusion from recent code augmentation studies~\cite{bielik2020adversarial,yu2022data,dong2025boosting} which states code augmentation has limited advantage in boosting the training of code models and also indicates that data augmentation for code is a challenging problem. 
\begin{table*}[h]
\centering
\caption{Effectiveness of data augmentation methods w.r.t. test accuracy $\uparrow$ (average $\pm$ standard deviation, \%) on original test data. \textbf{No Aug}: without data augmentation. The best results are highlighted in blue. Code understanding tasks include \textbf{Authorship attribution} (GCJ),  \textbf{Bug detection} (Refactory), \textbf{Problem classification} (Java250), and \textbf{Clone detection} (BigCloneBench).}
\label{tab:ACC_pre-trained}
\centering
\resizebox{0.95\columnwidth}{!}{
\begin{tabular}{clcccccc}
\cline{1-6}
\textbf{Model} & \textbf{DA method} & \textbf{GCJ}  & \textbf{Refactory} & \textbf{Java250} & \textbf{BigCloneBench} & \\ \cline{1-6} 
 & No Aug  & {90.96 ± 0.07} & {96.21 ± 0.13} & {96.36 ± 0.09} & {96.91 ± 0.12} & {} & \\ 
 &Refactor & {91.74 ± 0.11} & {95.95 ± 0.19} & {96.44 ± 0.12} & {96.97 ± 0.21} & {} & \\ 
 & SR & {70.66 ± 0.11} & {96.71 ± 0.11} & {96.21± 0.02} & {97.02 ± 0.15} & {} & \\ 
 & RI & {59.91 ± 0.21} & {96.43 ± 0.18} & {96.15 ± 0.06} & {97.38 ± 0.23} & {} & \\  
 & RS & {93.24 ± 0.13}  & {96.73 ± 0.15} & {96.49 ± 0.04} & {96.92 ± 0.12} & {} & \\  
  & RD & {76.98 ± 0.09}  & {96.46 ± 0.18} & {96.61 ± 0.12} & {97.11 ± 0.19} & {} & \\  
 & BT & {82.74 ± 0.15} & {94.31 ± 0.38} & {96.06 ± 0.06} & {97.13 ± 0.17} & {} & \\  
 & MixCode & {92.21 ± 0.25}& {96.76 ± 0.31} & {96.62 ± 0.23} & {-} & {} & \\  
\multirow{-9}{*}{CodeBERT} &  GenCode & \cellcolor[HTML]{DBE7FC}{96.23 ± 0.18} & \cellcolor[HTML]{DBE7FC}{99.32 ± 0.14} & \cellcolor[HTML]{DBE7FC}{98.89 ± 0.09} & \cellcolor[HTML]{DBE7FC}{98.93 ± 0.11} & {} & \\    \cline{1-6} 
 & No Aug & 93.96 ± 0.15 & {96.84 ± 0.16} & {96.41 ± 0.11} & {96.88 ± 0.09} & {} & \\ 
 &Refactor & {91.72 ± 0.21} & {95.49 ± 0.21} & {96.53 ± 0.15} & {97.11 ± 0.23} & {} & \\ 
 & SR & {89.46 ± 0.13} & {95.57 ± 0.13} & {96.46 ± 0.14} & {97.14 ± 0.16} & {} & \\ 
 & RI & {80.47 ± 0.21} & {94.51 ± 0.25} & {96.52 ± 0.09} & {97.35 ± 0.17} & {} & \\  
 & RS & {94.76 ± 0.15}  & {97.88 ± 0.15} & {96.44 ± 0.14} & {97.54 ± 0.12} & {} & \\  
  & RD & {91.71 ± 0.26}  & {96.53 ± 0.19} & {96.56 ± 0.21} & {97.16 ± 0.21} & {} & \\  
 & BT & {80.47 ± 0.05} & {95.99 ± 0.27} & {96.39 ± 0.13} & {97.11 ± 0.16} & {} & \\  
 & MixCode & {94.19 ± 0.25}& {97.29 ± 0.15} & {96.89 ± 0.19} & {-} & {} & \\  
\multirow{-9}{*}{GraphCodeBERT} & GenCode & \cellcolor[HTML]{DBE7FC}{97.93 ± 0.14}& \cellcolor[HTML]{DBE7FC}{99.45 ± 0.17} & \cellcolor[HTML]{DBE7FC}{99.16 ± 0.13} & \cellcolor[HTML]{DBE7FC}{99.23 ± 0.08} & {} & \\   \cline{1-6}
 & No Aug & 95.02 ± 0.07 & {97.22 ± 0.17} & {96.98 ± 0.08} & {97.38 ± 0.13} & {} & \\ 
 &Refactor & {92.21 ± 0.14} & {96.43 ± 0.22} & {96.92 ± 0.11} & {97.24 ± 0.12} & {} & \\ 
 & SR & {90.14 ± 0.21} & {94.67 ± 0.14} & {96.53 ± 0.05} & {97.31 ± 0.08} & {} & \\ 
 & RI & {78.77 ± 0.05} & {95.02 ± 0.15} & {96.11 ± 0.09} & {97.46 ± 0.13} & {} & \\  
 & RS & { 95.66 ± 0.14 }  & {97.97 ± 0.29} & {96.34 ± 0.11} & {97.03 ± 0.15} & {} & \\  
  & RD & {92.01 ± 0.11}  & {97.52 ± 0.26} & {97.21 ± 0.07} & {97.43 ± 0.07} & {} & \\  
 & BT & {81.45 ± 0.13} & {95.38 ± 0.11} & {96.45 ± 0.14} & {97.65 ± 0.14} & {} & \\  
 & MixCode & {94.32 ± 0.16}& {97.32 ± 0.13} & {96.91 ± 0.11} & {-} & {} & \\  
\multirow{-9}{*}{CodeT5} & GenCode & \cellcolor[HTML]{DBE7FC}{98.85 ± 0.11} & \cellcolor[HTML]{DBE7FC}{99.64} ± 0.06 & \cellcolor[HTML]{DBE7FC}{99.38 ± 0.11} & \cellcolor[HTML]{DBE7FC}{99.34 ± 0.12} &  & \\ \cline{1-6}

 & No Aug & 98.91 ± 0.02 & {98.12 ± 0.01} & {97.83 ± 0.02} & {98.75 ± 0.03} & {} & \\ 
 &Refactor & {96.92 ± 0.01} & {97.99 ± 0.03} & {97.25 ± 0.02} & {98.61 ± 0.01} & {} & \\ 
 & SR & {92.36 ± 0.04} & {97.27 ± 0.04} & {97.27 ± 0.03} & {98.46 ± 0.04} & {} & \\ 
 & RI & {90.47 ± 0.05} & {96.82 ± 0.03} & {97.26 ± 0.02} & {98.31 ± 0.02} & {} & \\  
 & RS & {97.66 ± 0.06}  & {97.41 ± 0.06} & {98.19 ± 0.01} & {98.15 ± 0.01} & {} & \\  
  & RD & {98.33 ± 0.03}  & {97.02 ± 0.02} & {97.72 ± 0.02} & {98.36 ± 0.03} & {} & \\  
 & BT & {89.92 ± 0.02} & {97.09 ± 0.05} & {97.41 ± 0.02} & {98.12 ± 0.05} & {} & \\  
 & MixCode & {98.56 ± 0.04}& {98.21 ± 0.08} & {98.23 ± 0.03} & {-} & {} & \\  
\multirow{-9}{*}{StarCoder2-7B} & GenCode & \cellcolor[HTML]{DBE7FC}{98.93 ± 0.04}& \cellcolor[HTML]{DBE7FC}{99.65 ± 0.02} & \cellcolor[HTML]{DBE7FC}{99.41 ± 0.05} & \cellcolor[HTML]{DBE7FC}{99.36 ± 0.09} & {} & \\   \cline{1-6}

 & No Aug & 98.95 ± 0.03 & 98.19 ± 0.03 & {98.11 ± 0.02} & {98.78 ± 0.01} & {} & \\ 
 &Refactor & {97.22 ± 0.03} & {97.51 ± 0.02} & {97.31 ± 0.01} & {98.32 ± 0.04} & {} & \\ 
 & SR & {92.45 ± 0.02} & {97.47 ± 0.03} & {97.37 ± 0.04} & {98.51 ± 0.05} & {} & \\ 
 & RI & {90.72 ± 0.03} & {96.31 ± 0.02} & {98.07 ± 0.05} & {98.34 ± 0.04} & {} & \\  
 & RS & {96.89 ± 0.05}  & {97.52 ± 0.05} & {97.31 ± 0.02} & {98.02 ± 0.06} & {} & \\  
  & RD & {96.83 ± 0.01}  & {96.89 ± 0.04} & {97.52 ± 0.03} & {98.25 ± 0.04} & {} & \\  
 & BT & {91.07 ± 0.06} & {97.15 ± 0.03} & {98.02 ± 0.02} & {98.06 ± 0.08} & {} & \\  
 & MixCode & {98.61 ± 0.03}& {98.46 ± 0.06} & {98.53 ± 0.04} & {-} & {} & \\  
\multirow{-9}{*}{Qwen2.5-Coder-7B} & GenCode & \cellcolor[HTML]{DBE7FC}{98.96 ± 0.04}& \cellcolor[HTML]{DBE7FC}{99.65 ± 0.01} & \cellcolor[HTML]{DBE7FC}{99.42 ± 0.04} & \cellcolor[HTML]{DBE7FC}{99.36 ± 0.05} & {} & \\   \cline{1-6}
\end{tabular}
}

\vspace{-3mm}
\end{table*}

Then, compared to the existing data augmentation methods, the results demonstrate that GenCode consistently enhances the training and achieves the best results across all the considered datasets and models. Concretely, on average, for pre-trained code models, GenCode generates models with 4.52\%, 3.05\%, 2.56\%, and 2.11\% higher accuracy than \emph{No Aug} in GCJ, Refactory, Java250, and BigCloneBench, respectively. Besides, compared to MixCode, GenCode can achieve 4.1\%, 2.34\%, and 2.33\% better results in GCJ, Refactory, and Java250 datasets, respectively. Moving to LLMs, GenCode improves the performance of StarCoder2 across all tasks by up to 0.93\% on average, and achieves a 0.84\% average improvement for Qwen2.5-Coder.

\begin{figure*}[!tb]
\centering
\subfigure[Authorship attribution]{
\begin{minipage}[t]{0.5\linewidth}
\includegraphics[width=1.0\linewidth]{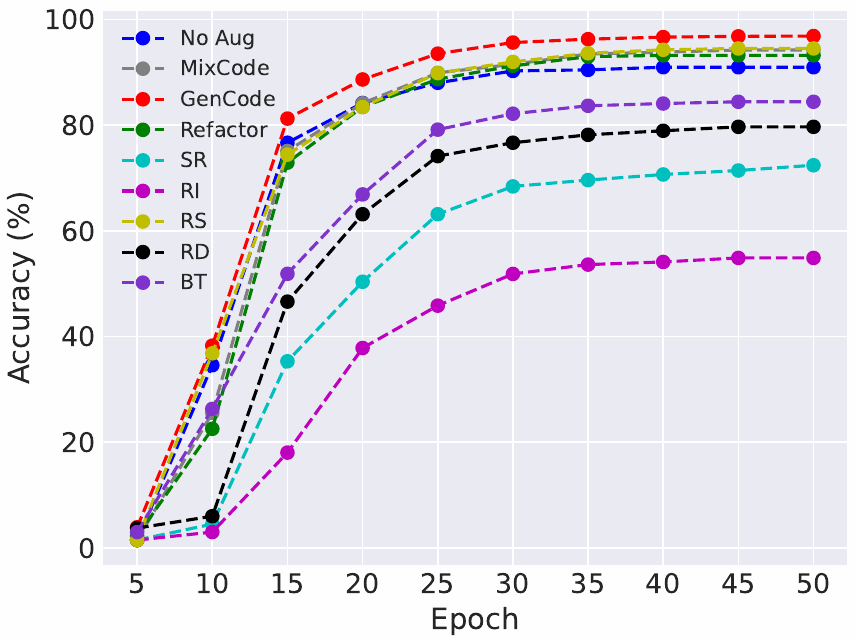}
\end{minipage}%
}%
\subfigure[Bug detection]{
\begin{minipage}[t]{0.5\linewidth}
\includegraphics[width=1.0\linewidth]{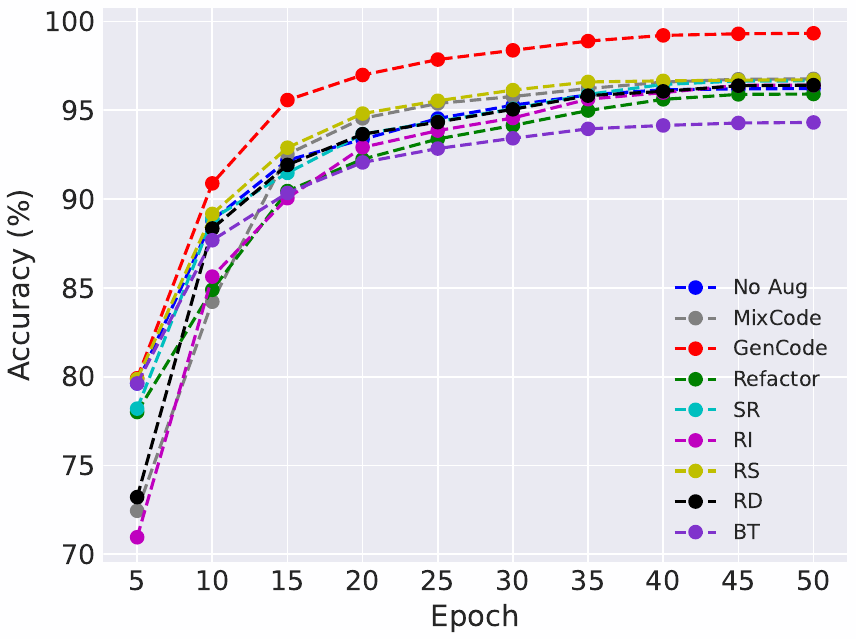}
\end{minipage}%
}%

\subfigure[Problem classification]{
\begin{minipage}[t]{0.5\linewidth}
\centering
\includegraphics[width=1.0\linewidth]{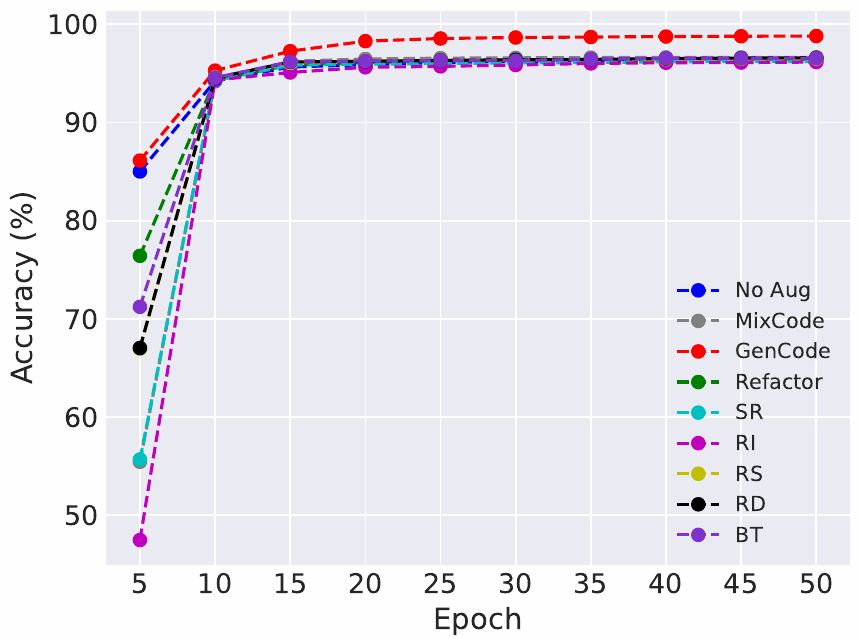}
\end{minipage}%
}%
\subfigure[Clone detection]{
\begin{minipage}[t]{0.5\linewidth}
\centering
\includegraphics[width=1.0\linewidth]{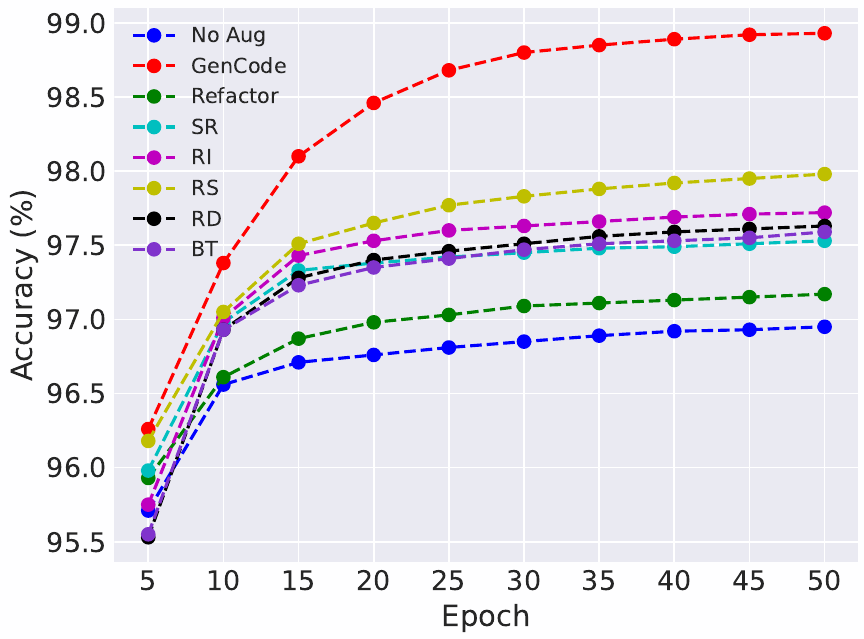}
\end{minipage}%
}%
\caption{Convergence speed of CodeBERT using different code augmentation methods in each task.}
\label{fig:log}
\end{figure*}

\textbf{Convergence speed analysis.} Fig.~\ref{fig:log} depicts the average test accuracy at each training epoch of CodeBERT using different data augmentation methods. The results demonstrated that GenCode consistently outperforms other baseline methods during the whole training process. We can conclude that the choice of the number of training epochs will not affect the advantages of our method over other methods.   

\textbf{Visualizing principal component analysis.} Furthermore, we explore why models trained by GenCode have higher accuracy than models trained using other methods by analyzing the code representation capability of trained code embeddings. Specifically, 1) first, we extract the code embeddings produced by each trained model using all the test data as code features. 2) Then, since the extracted features are high dimensions and hard to analyze, we reduce the dimensions of code data features using the \emph{Principal Component Analysis~(PCA)}~\cite{mackiewicz1993principal} algorithm. 3) After that, we visualize the reduced 2-dimensional vectors. Normally, the features extracted by a well-trained model are more clearly distributed based on the data labels. 
\begin{figure*}[htbp]
\centering
\subfigure[Refactor]{
\begin{minipage}[t]{0.45\linewidth}
\centering
\includegraphics[width=1.0\linewidth]{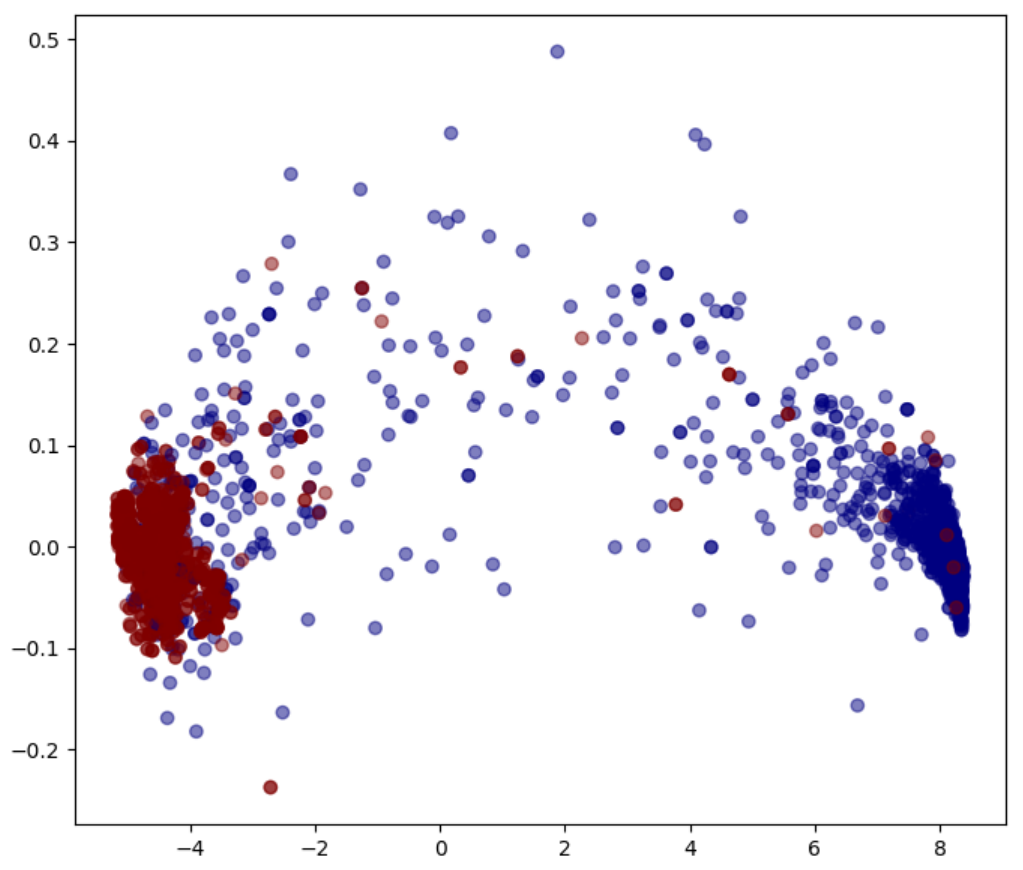}
\end{minipage}%
}%
\subfigure[SR]{
\begin{minipage}[t]{0.45\linewidth}
\centering
\includegraphics[width=1.0\linewidth]{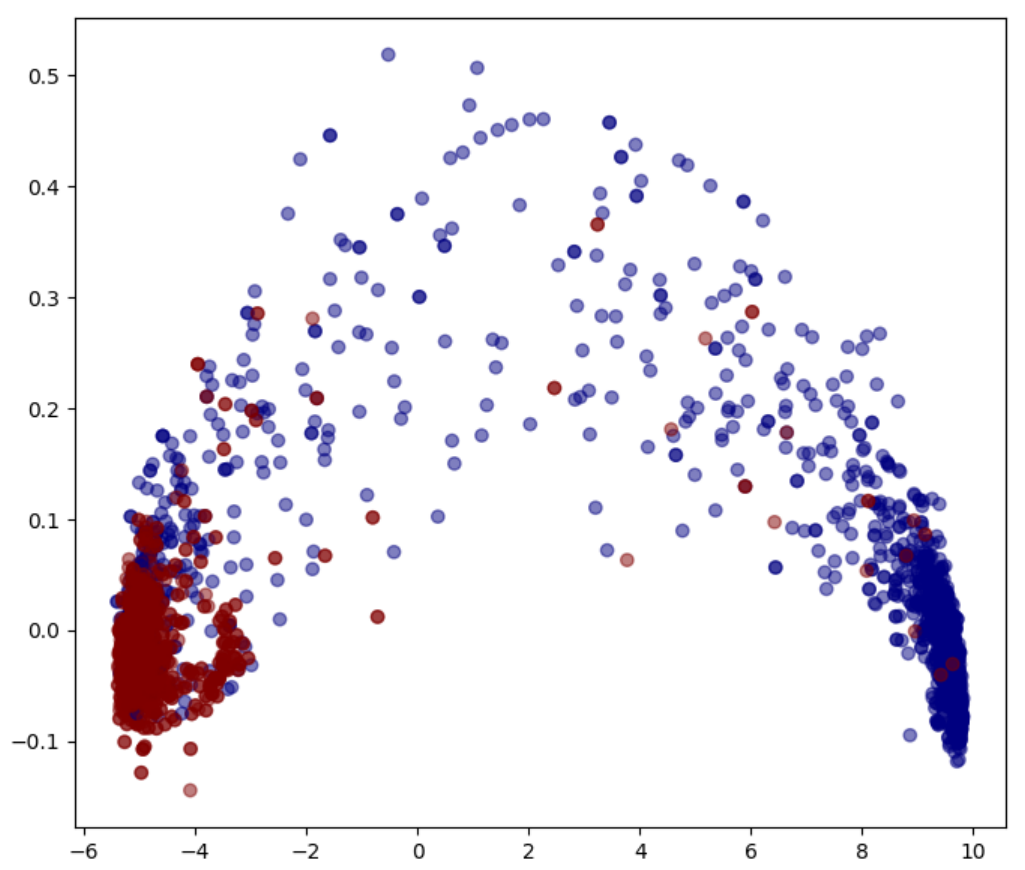}
\end{minipage}%
}%

\centering
\subfigure[RI]{
\begin{minipage}[t]{0.45\linewidth}
\centering
\includegraphics[width=1.0\linewidth]{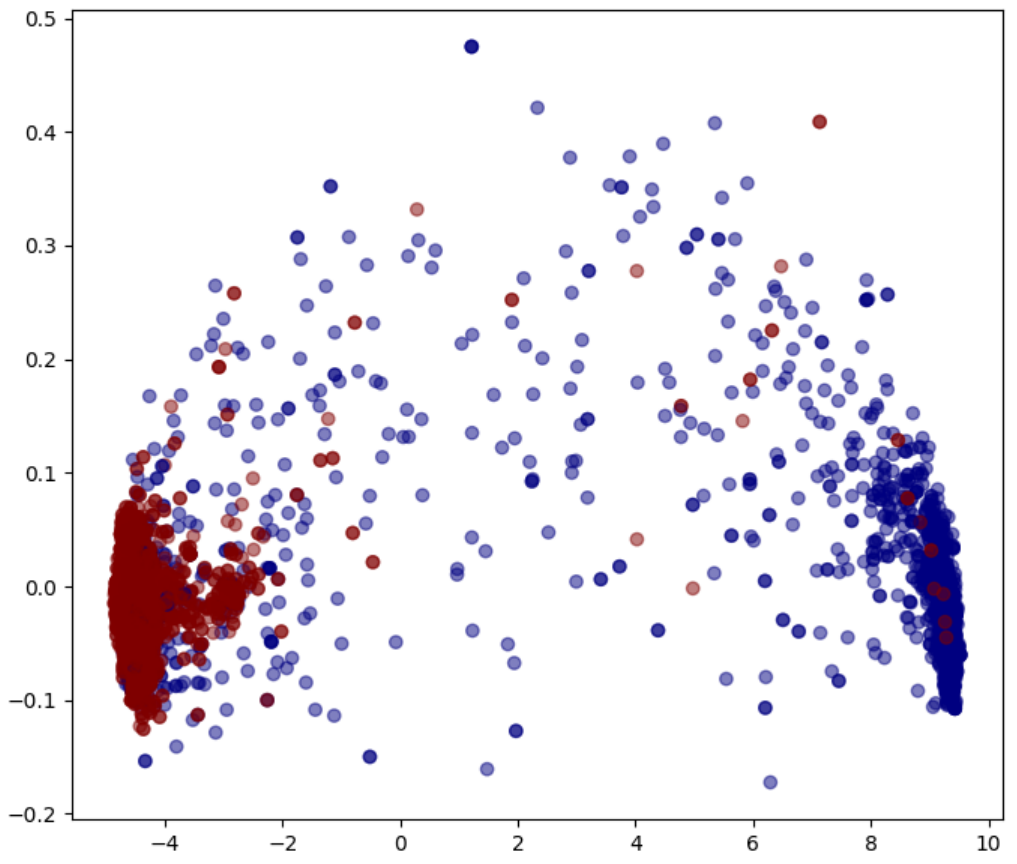}
\end{minipage}%
}%
\centering
\subfigure[RS]{
\begin{minipage}[t]{0.45\linewidth}
\centering
\includegraphics[width=1.0\linewidth]{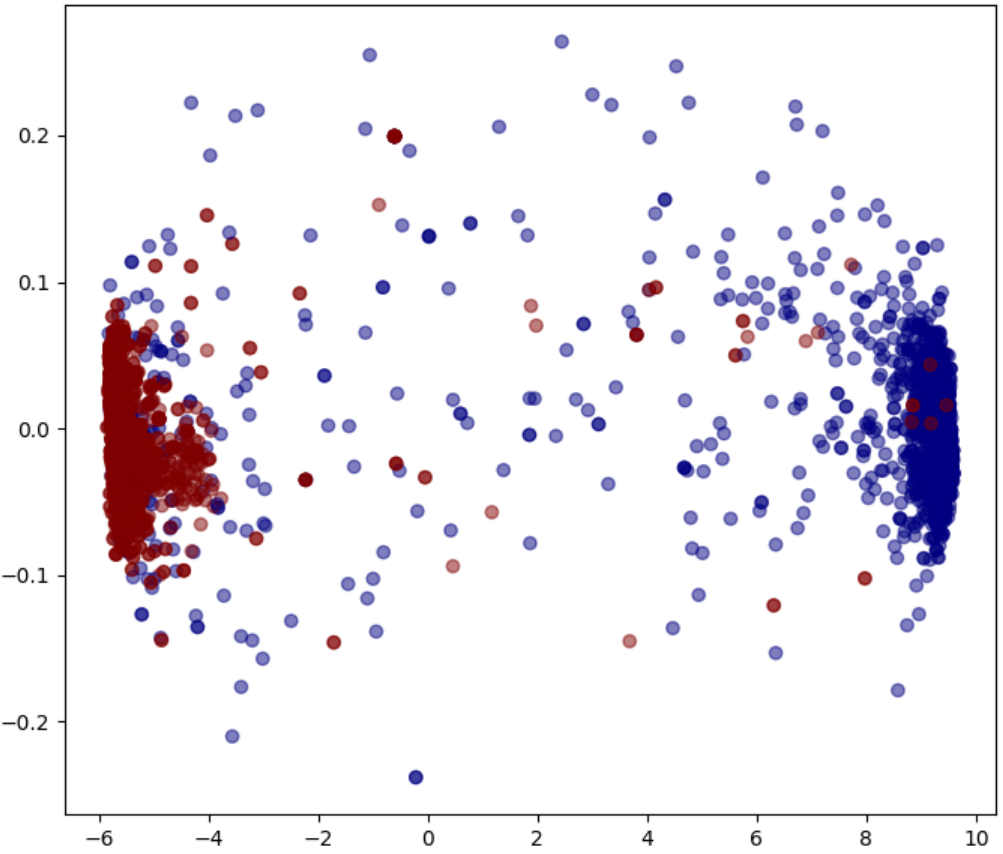}
\end{minipage}%
}%

\centering
\subfigure[RD]{
\begin{minipage}[t]{0.45\linewidth}
\centering
\includegraphics[width=1.0\linewidth]{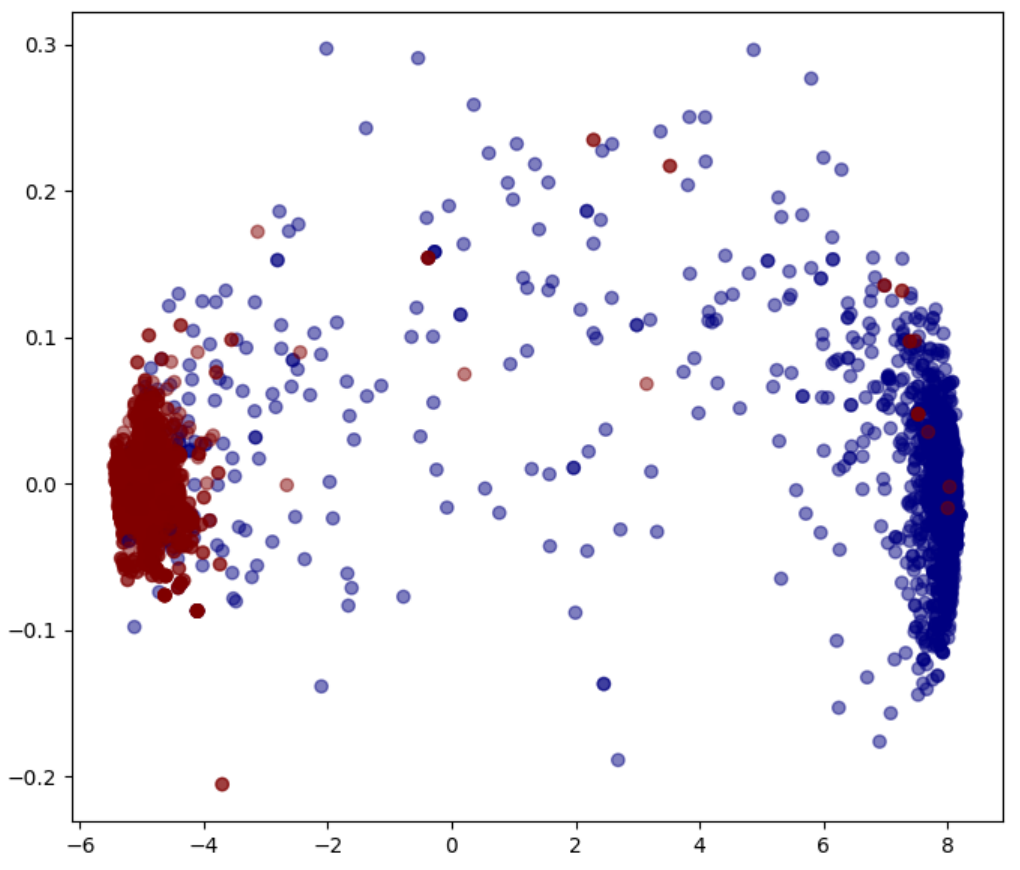}
\end{minipage}%
}%
\centering
\subfigure[BT]{
\begin{minipage}[t]{0.45\linewidth}
\centering
\includegraphics[width=1.0\linewidth]{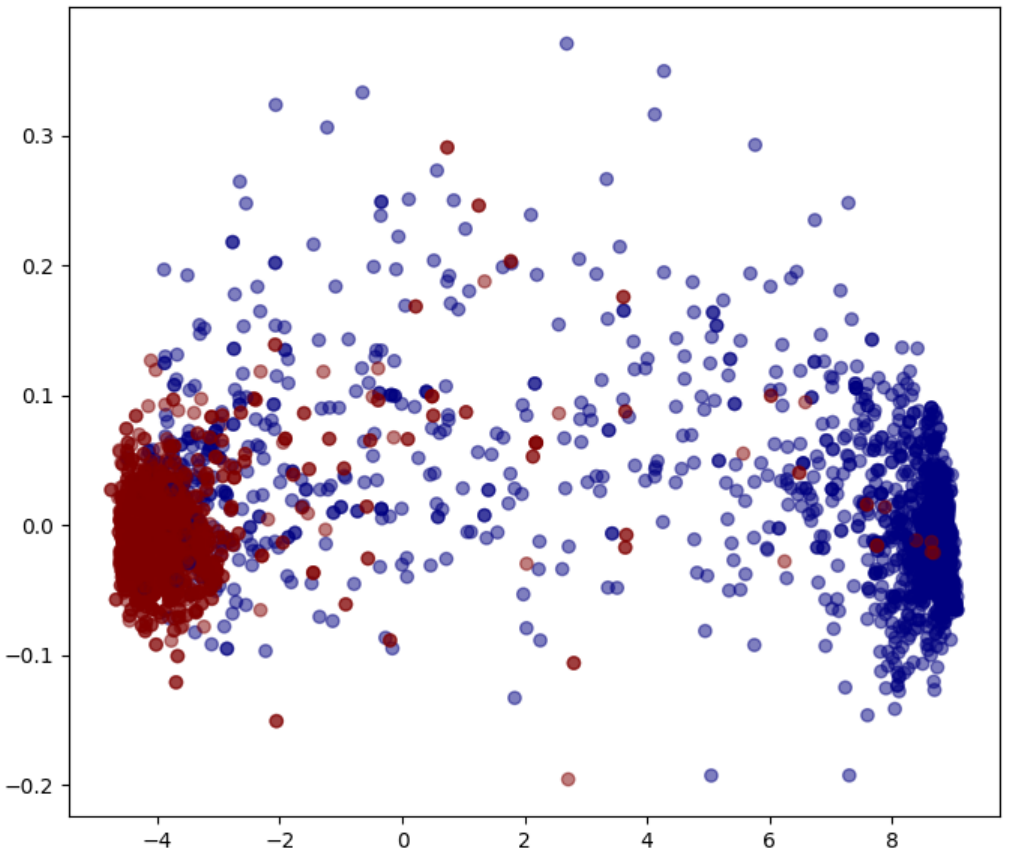}
\end{minipage}%
}%

\centering
\subfigure[MixCode]{
\begin{minipage}[t]{0.45\linewidth}
\centering
\includegraphics[width=1.0\linewidth]{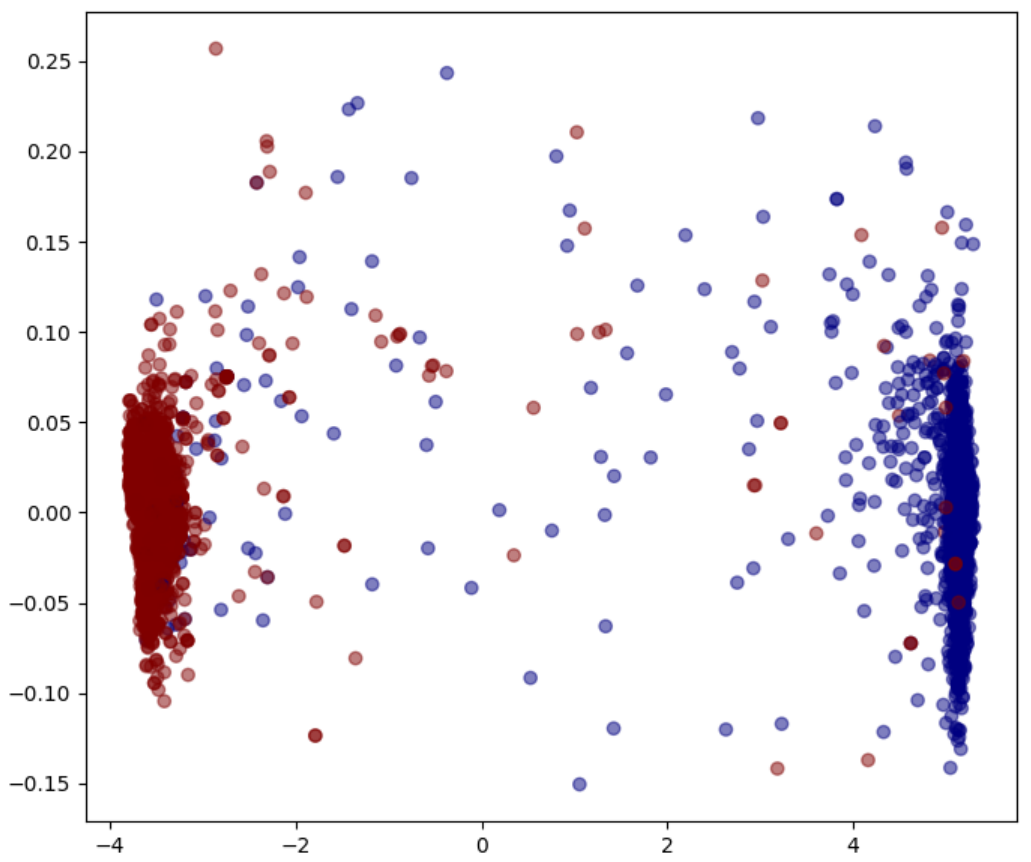}
\end{minipage}%
}%
\centering
\subfigure[GenCode]{
\begin{minipage}[t]{0.45\linewidth}
\centering
\includegraphics[width=1.0\linewidth]{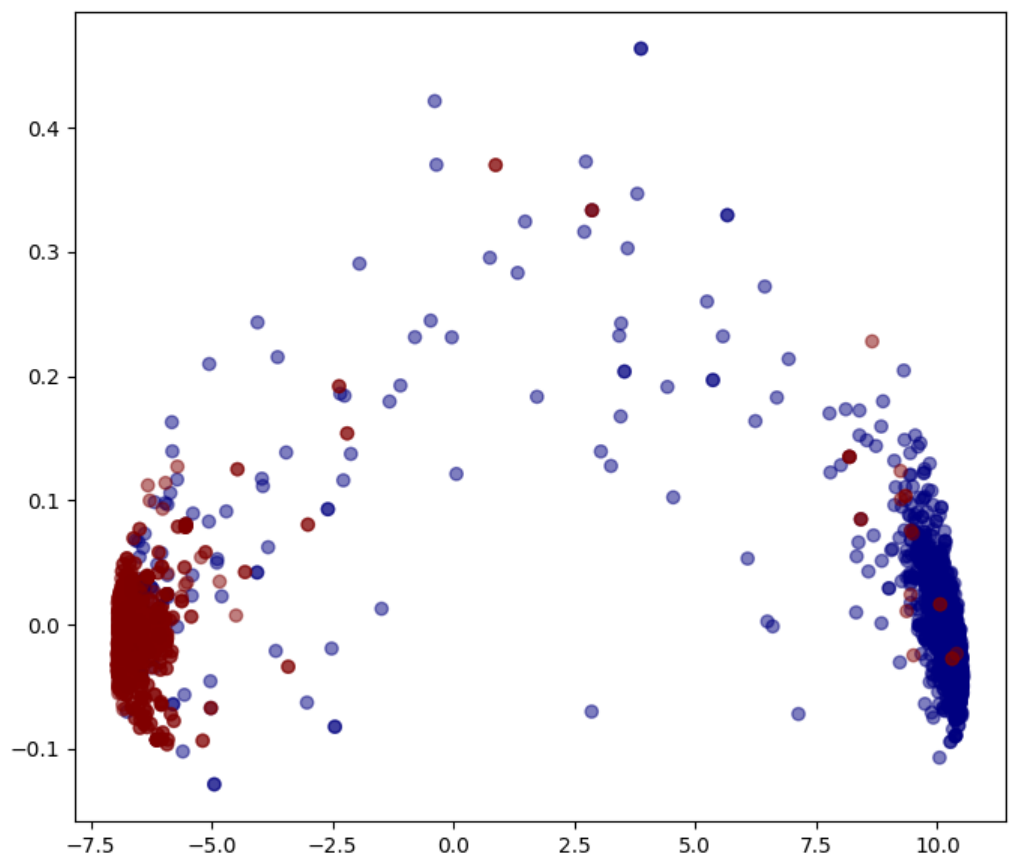}
\end{minipage}%
}%
\caption{Visualization of code embeddings after dimension reduction using Principal Component Analysis (PCA). Model: CodeBERT, dataset: Refactory, task: Bug detection. 
}
\label{fig: log_clone}
\end{figure*}
Fig.~\ref{fig: log_clone} depicts the visualized features extracted from CodeBERT using the Refactory dataset. We can see that the code embeddings trained by using GenCode are more clearly distributed than embedding extracted by other models, i.e., the buggy code and correct code are greatly different in the figure. These results indicate that GenCode can train good code embeddings and, therefore, produce better accurate code models for code understanding tasks. Besides, we computed the average distance between the data points from different classes, i.e., the average distance between red points and blue points in Fig.~\ref{fig: log_clone}. The results from Table~\ref{tab:distance} confirm that GenCode can better distinguish the two classes. Interestingly, we found that MixCode has the smallest distance result. That means even though MixCode can train code models with good performance, the code models have low confidence in their predictions, thus, the data are near the decision boundaries. This phenomenon is similar to the findings in~\cite{verma2019manifold}, where interpolating hidden state representations yields two beneficial properties: 1) class representations are flattened into a minimal number of directions of variation, and 2) all points between these flat representations are assigned low-confidence predictions.

\begin{table}[!tb]
\caption{ Average distance between the data points from different classes. The best is highlighted with a blue background, and the worst is highlighted with a gray background. }
\label{tab:distance}
\centering
\resizebox{0.75\columnwidth}{!}{
\begin{tabular}{clcccccccc}
 \cline{1-6}
\textbf{} & \textbf{DA method} & \textbf{Refactor}  & \textbf{SR} & \textbf{RI} & \textbf{RS} & \\ 
 & \textbf{Average distance}  & {6.02} & {8.05} & {7.02} & {8.06} & {} & \\   \cline{1-6}  
 \textbf{} & \textbf{DA method} &  \textbf{RD} & \textbf{BT} & \textbf{MixCode} & \textbf{GenCode}\\ 
 & \textbf{Average distance}  & {7.03} & {6.00} & \cellcolor[HTML]{C0C0C0}{4.05} & \cellcolor[HTML]{DBE7FC}{8.78
} & {} & \\  
\cline{1-6} 
\end{tabular}
}
\vspace{-3mm}
\end{table}

\begin{table}[h]
\centering
    \setlength{\tabcolsep}{3pt} 
    \renewcommand{\arraystretch}{1.3}
    \tiny
\caption{Results of statistical tests on Accuracy \& Natural Robustness.  A gray background highlights that the comparison is statistically significant. Statistical tests method: \emph{Wilcoxon signed-rank test}. Across CodeBERT, GraphCodeBERT, CodeT5, StarCoder2, and Qwen2.5-Coder models.}
\label{table:statistical_testing_acc_robust}
\centering
\resizebox{\textwidth}{!}{
\begin{tabular}{lcccccccccccccccc}
\hline
                    & DA method & No Aug & Refactor & SR & RI &  RS &  RD & BT & MixCode \\ \hline
\multirow{1}{*}{{\begin{tabular}[c]{@{}c@{}}Accuracy \end{tabular}}} & GenCode &\cellcolor[HTML]{C0C0C0}{\color[HTML]{000000} 1.907e-06 }  & \cellcolor[HTML]{C0C0C0}{\color[HTML]{000000}1.907e-06 }  & \cellcolor[HTML]{C0C0C0}{\color[HTML]{000000}1.907e-06 }  & \cellcolor[HTML]{C0C0C0}{\color[HTML]{000000}8.845e-05 }  &  \cellcolor[HTML]{C0C0C0}{\color[HTML]{000000}1.907e-06 }  &  \cellcolor[HTML]{C0C0C0}{\color[HTML]{000000}1.907e-06 }  & \cellcolor[HTML]{C0C0C0}{\color[HTML]{000000}1.907e-06 }  & \cellcolor[HTML]{C0C0C0}{\color[HTML]{000000}0.001 }                             \\
\hline                  
\multirow{1}{*}{{\begin{tabular}[c]{@{}c@{}} Natural Robustness \end{tabular}}} & GenCode &\cellcolor[HTML]{C0C0C0}{\color[HTML]{000000}1.907e-06 }  & {- }  & \cellcolor[HTML]{C0C0C0}{\color[HTML]{000000}1.907e-06 }  & \cellcolor[HTML]{C0C0C0}{\color[HTML]{000000}1.907e-06 }  &  \cellcolor[HTML]{C0C0C0}{\color[HTML]{000000}1.907e-06 }  &  \cellcolor[HTML]{C0C0C0}{\color[HTML]{000000}1.316e-04 }  & \cellcolor[HTML]{C0C0C0}{\color[HTML]{000000}1.907e-06 }  & \cellcolor[HTML]{C0C0C0}{\color[HTML]{000000}6.533e-04 }    \\

\hline  

\end{tabular}
}
\vspace{-3mm}
\end{table}
\textbf{Statistical analysis.} We conducted statistical tests with \emph{Wilcoxon signed-rank test}, across all code models. The results from the upper of Table~\ref{table:statistical_testing_acc_robust} show that GenCode significantly (with \emph{p}-value $<$ 0.05) outperforms other methods including MixCode. In conclusion, GenCode is effective in training accurate code models.

\begin{tcolorbox}
\textbf{Answer to RQ1}: GenCode consistently outperforms baselines in training more accurate code understanding models. Specifically, for pre-trained code models, GenCode yields an average accuracy improvement of 3.06\% over models trained without data augmentation and 2.92\% over those trained with the SOTA MixCode method. For LLMs, GenCode achieves average improvements of 0.93\% for StarCoder2 and 0.84\% for Qwen2.5-Coder.
\end{tcolorbox}

\subsection{RQ2: How effective is GenCode in producing robust code models?}
\label{sec:results_robustness}

 \textbf{Adversarial robustness evaluation.} 
 \begin{table}[h]
\caption{Effectiveness of data augmentation methods w.r.t. ASR $\downarrow$ (\%) on test data. The best results are highlighted in blue. Code understanding tasks include \textbf{Authorship attribution} (GCJ),  \textbf{Bug detection} (Refactory), \textbf{Problem classification} (Java250), and \textbf{Clone detection} (BigCloneBench).}
\label{tab:Robustness_pre-trained}
\centering
\resizebox{0.95\columnwidth}{!}{
\begin{tabular}{clcccccccccccc}
\hline
 & \multicolumn{1}{c}{} & \multicolumn{2}{c}{\textbf{GCJ}} & \multicolumn{2}{c}{\textbf{Refactory}} & \multicolumn{2}{c}{\textbf{Java250}} & \multicolumn{2}{c}{\textbf{BigCloneBench}} \\
\multirow{-2}{*}{\textbf{Model}} & \multicolumn{1}{l}{\multirow{-2}{*}{\textbf{DA method}}} & \textbf{MHM} & \textbf{ALERT} & \textbf{\textbf{MHM}} & \textbf{ALERT} & \textbf{MHM} & \textbf{ALERT} & \textbf{MHM} & \textbf{ALERT}  \\ \hline
 & No Aug & \multicolumn{1}{c}{{33.31}} & {56.13} & \multicolumn{1}{c}{{36.31 }} & {42.31} & \multicolumn{1}{c}{40.55} & {53.37} & \multicolumn{1}{c}{9.74} & {24.61} \\ 
 & Refactor & \multicolumn{1}{c}{{35.56}} & {52.84} & \multicolumn{1}{c}{{32.64}} & {40.89} & \multicolumn{1}{c}{38.35} & {52.56} & \multicolumn{1}{c}{8.55} & {18.29} \\ 
 & SR & \multicolumn{1}{c}{{35.46}} & {52.71} & \multicolumn{1}{c}{{39.25}} & {53.91} & \multicolumn{1}{c}{42.01} & {53.79} & \multicolumn{1}{c}{12.52} & {26.11} \\  
 & RI & \multicolumn{1}{c}{{40.78}} & {65.76} & \multicolumn{1}{c}{{38.45}} & \multicolumn{1}{c}{{45.57}} & {43.15} & \multicolumn{1}{c}{55.07} & {8.31} & \multicolumn{1}{c}{30.08} & { }\\ 
 & RS & \multicolumn{1}{c}{{28.26}} & {52.44} & \multicolumn{1}{c}{{38.63}} & {45.98} & \multicolumn{1}{c}{39.36} & {50.79} & \multicolumn{1}{c}{13.41} & {30.13}\\ 
 & RD & \multicolumn{1}{c}{{35.65}} & {55.46} & \multicolumn{1}{c}{{30.23}} & {41.32} & \multicolumn{1}{c}{38.31} & {52.42} & \multicolumn{1}{c}{9.58} & {30.15} \\ 
 & BT & \multicolumn{1}{c}{{40.38 }} & {55.95} & \multicolumn{1}{c}{62.98} & 69.87 & \multicolumn{1}{c}{42.21} & {53.32} & \multicolumn{1}{c}{11.29} & {26.32}  \\ 
 & MixCode & \multicolumn{1}{c}{{29.28 }} & {52.53} & \multicolumn{1}{c}{{31.04}} & {41.01} & \multicolumn{1}{c}{36.14} & {50.67} & \multicolumn{1}{c}{-} & {-}\\ 
\multirow{-9}{*}{CodeBERT} & GenCode &  \multicolumn{1}{c}{\cellcolor[HTML]{DBE7FC}22.33} & \cellcolor[HTML]{DBE7FC}{46.62} &  \multicolumn{1}{c}{\cellcolor[HTML]{DBE7FC}23.75} & \cellcolor[HTML]{DBE7FC}{31.25} & \cellcolor[HTML]{DBE7FC}29.89 & \cellcolor[HTML]{DBE7FC}40.65 & \cellcolor[HTML]{DBE7FC}7.12 & \cellcolor[HTML]{DBE7FC}15.83 \\ \hline
 & No Aug & \multicolumn{1}{c}{{31.23}} & {56.02} & \multicolumn{1}{c}{{22.97}} & {30.26} & \multicolumn{1}{c}{23.16} & {42.66} & \multicolumn{1}{c}{4.19} & {12.22} \\ 
 & Refactor & \multicolumn{1}{c}{{30.31}} & {56.58} & \multicolumn{1}{c}{{24.26}} & {32.57} & \multicolumn{1}{c}{23.13} & {42.42} & \multicolumn{1}{c}{6.05} & {6.17} \\ 
 & SR & \multicolumn{1}{c}{{40.02}} & {66.68} & \multicolumn{1}{c}{{28.39}} & {47.15} & \multicolumn{1}{c}{22.92} & {42.18} & \multicolumn{1}{c}{6.13} & {10.16} \\  
 & RI & \multicolumn{1}{c}{{43.51}} & {65.75} & \multicolumn{1}{c}{{25.08}} & \multicolumn{1}{c}{{32.87}} & {22.37} & \multicolumn{1}{c}{42.33} & {2.23} & \multicolumn{1}{c}{6.28} \\ 
 & RS & \multicolumn{1}{c}{{30.18}} & {54.75} & \multicolumn{1}{c}{{21.85}} & {29.67} & \multicolumn{1}{c}{22.68} & {41.98} & \multicolumn{1}{c}{6.03} & {11.94}\\ 
 & RD & \multicolumn{1}{c}{{31.73}} & { 60.92 } & \multicolumn{1}{c}{{20.35}} & {26.98} & \multicolumn{1}{c}{22.39} & {41.84} & \multicolumn{1}{c}{6.18} & {8.27} \\ 
 & BT & \multicolumn{1}{c}{{41.65}} & {62.99} & \multicolumn{1}{c}{{35.56}} & {46.23} & \multicolumn{1}{c}{22.91} & {42.32} & \multicolumn{1}{c}{8.18} & {14.12}  \\ 
 & MixCode & \multicolumn{1}{c}{{30.26}} & {54.49} & \multicolumn{1}{c}{\cellcolor[HTML]{DBE7FC}{14.89}} & {26.53} & \multicolumn{1}{c}{20.03} & {42.01} & \multicolumn{1}{c}{-} & {-}\\ 
\multirow{-9}{*}{GraphCodeBERT} & GenCode & \multicolumn{1}{c}{\cellcolor[HTML]{DBE7FC}{26.43}} & \cellcolor[HTML]{DBE7FC}{48.87} & \multicolumn{1}{c}{{15.78}} & \cellcolor[HTML]{DBE7FC}{21.92} & \cellcolor[HTML]{DBE7FC}{18.76} & \cellcolor[HTML]{DBE7FC}{33.81} & \cellcolor[HTML]{DBE7FC}{1.86} & \cellcolor[HTML]{DBE7FC}{4.62} \\ \hline
 & No Aug & \multicolumn{1}{c}{{30.78}} & {55.67} & \multicolumn{1}{c}{{22.45}} & {29.69} & \multicolumn{1}{c}{31.76} & {49.82} & \multicolumn{1}{c}{10.32} & {22.97} \\ 
 & Refactor & \multicolumn{1}{c}{{30.12}} & {55.93} & \multicolumn{1}{c}{{28.76}} & {35.45} & \multicolumn{1}{c}{30.66} & {47.12} & \multicolumn{1}{c}{9.32} & {22.05} \\ 
 & SR & \multicolumn{1}{c}{{39.21}} & {51.84} & \multicolumn{1}{c}{{28.24}} & {36.26} & \multicolumn{1}{c}{28.65} & {46.01} & \multicolumn{1}{c}{15.87} & {25.98} \\  
 & RI & \multicolumn{1}{c}{{41.02}} & { 63.46 } & \multicolumn{1}{c}{{24.78}} & \multicolumn{1}{c}{{33.61}} & {27.14} & \multicolumn{1}{c}{46.76} & {10.43} & \multicolumn{1}{c}{29.76} \\ 
 & RS & \multicolumn{1}{c}{{29.35}} & {51.74} & \multicolumn{1}{c}{{21.32}} & {27.76} & \multicolumn{1}{c}{28.45} & {48.45} & \multicolumn{1}{c}{14.33} & {30.43}\\ 
 & RD & \multicolumn{1}{c}{{31.04}} & { 56.85 } & \multicolumn{1}{c}{{19.93}} & {24.87} & \multicolumn{1}{c}{26.09} & {45.24} & \multicolumn{1}{c}{9.98} & {28.87} \\ 
 & BT & \multicolumn{1}{c}{{39.65}} & {54.24} & \multicolumn{1}{c}{{33.62}} & {43.67} & \multicolumn{1}{c}{29.47} & {45.38} & \multicolumn{1}{c}{10.43} & {25.34}  \\ 
 & MixCode & \multicolumn{1}{c}{{29.82}} & \cellcolor[HTML]{DBE7FC}{42.56} & \multicolumn{1}{c}{{12.49}} & \cellcolor[HTML]{DBE7FC}{18.32} & \multicolumn{1}{c}{28.73} & {48.87} & \multicolumn{1}{c}{-} & {-}\\ 
\multirow{-9}{*}{CodeT5} & GenCode & \multicolumn{1}{c}{\cellcolor[HTML]{DBE7FC}{23.57}} & {44.76} & \multicolumn{1}{c}{\cellcolor[HTML]{DBE7FC}{12.13}} & {19.56} & \cellcolor[HTML]{DBE7FC}{21.63} & \cellcolor[HTML]{DBE7FC}{39.76} & \cellcolor[HTML]{DBE7FC}{7.39} & \cellcolor[HTML]{DBE7FC}{12.15} \\ \hline
\end{tabular}
}
\vspace{-3mm}
\end{table}
Column~\emph{MHM} and~\emph{ALERT} in Table~\ref{tab:Robustness_pre-trained} summarize the attack success rate of adversarial attack methods against code models. First, we can see that similar to the observation in Accuracy Evaluation, not all data augmentation methods can improve the robustness of code models. This finding is consistent with the existing works~\cite{dong2025boosting}.  Specifically, in the baseline methods, only \textit{RS} and MixCode have a positive impact on the robustness of code models. Considering our method, GenCode consistently improves the robustness of code models and achieves the best results in 21 out of 24 cases. On average, compared to \emph{No Aug}, models trained by GenCode have an 8.42\% lower ASR~(higher robustness). This result breaks the previous finding~\cite{bielik2020adversarial} that states simply adding additional generated code to the training dataset to train code models has limited benefits in improving their robustness. On the other hand, compared to MixCode, GenCode can train code models with 4.90\% higher robustness on average. 

\textbf{Natural robustness evaluation.} Table~\ref{tab:natural_robust} presents the result of the natural robustness evaluation.  For a fair comparison,  we remove \emph{Refactor} from the column~\emph{DA method}. The results indicate that most text-oriented data augmentation methods~\cite{dong2025boosting}, such as \emph{BT} and \emph{RI}, do not contribute significantly to improving natural robustness. Sampling-based code augmentation methods, such as MixCode, yield limited improvements. For instance, in code multi-classification tasks including \emph{authorship attribution (GCJ)}, MixCode achieves an average improvement of only 0.56\% compared to training without data augmentation. Although GenCode outperforms all baselines in Table~\ref{tab:natural_robust}, the improvement remains limited, particularly for LLMs. For example, GenCode achieves an average improvement of only 0.98\% for StarCoder2-7B and 0.84\% for Qwen2.5-Coder-7B across all tasks. This observation aligns with findings from existing studies~\cite{dong2025boosting,ding2024data}, suggesting that the effectiveness of data augmentation diminishes as model scale increases. 
\begin{table*}[h]
\centering
\caption{Effectiveness of data augmentation methods w.r.t. test accuracy $\uparrow$ (average $\pm$ standard deviation, \%) on transformed test data. \textbf{No Aug}: without data augmentation. The best results are highlighted in blue. Code understanding tasks include \textbf{Authorship attribution} (GCJ),  \textbf{Bug detection} (Refactory), \textbf{Problem classification} (Java250), and \textbf{Clone detection} (BigCloneBench).}
\label{tab:natural_robust}
\centering
\resizebox{0.95\columnwidth}{!}{
\begin{tabular}{clcccccc}
\cline{1-6}
\textbf{Model} & \textbf{DA method} & \textbf{GCJ}  & \textbf{Refactory} & \textbf{Java250} & \textbf{BigCloneBench} & \\ \cline{1-6} 
 & No Aug  & {87.57 ± 0.13} & {92.12 ± 0.09} & {89.68 ± 0.16} & {91.61 ± 0.18} & {}  \\ 

 & SR & 62.78 ± 0.22 & {89.87 ± 0.15} & {85.51 ± 0.07} & {86.21 ± 0.12} & { } & {}  \\ 
 & RI & 50.35 ± 0.19 & {90.23 ± 0.19} & {85.21 ± 0.11} & {92.02 ± 0.13} & { } & {}  \\ 
 & RS & 88.23 ± 0.13 & {90.56 ± 0.12} & {89.97 ± 0.14} & {85.65 ± 0.11} & { } & {}  \\  
  & RD & 69.34 ± 0.18 & {93.78 ± 0.11} &{90.34 ± 0.13} & {91.87 ± 0.12} & { } & {}  \\   
 & BT & 75.43 ± 0.15 & {82.35 ± 0.21} & {85.23 ± 0.15} & {87.78 ± 0.09} & { } & {}  \\  
 & MixCode & {89.11 ± 0.13 }& {95.36 ± 0.09} & {93.19 ± 0.12} & {-} & {} & \\  
\multirow{-8}{*}{CodeBERT} &  GenCode & \cellcolor[HTML]{DBE7FC}{91.23 ± 0.14} & \cellcolor[HTML]{DBE7FC}{95.63 ± 0.14} & \cellcolor[HTML]{DBE7FC}{93.62 ± 0.09} & \cellcolor[HTML]{DBE7FC}{92.56 ± 0.15} & {} & \\    \cline{1-6} 
 
 & No Aug  & {87.86 ± 0.12} & {92.04 ± 0.03} & {90.58 ± 0.08} & {91.89 ± 0.09} & {}  \\ 

 & SR & 66.45 ± 0.09 & {89.12 ± 0.06} & {91.05 ± 0.12} & {89.23 ± 0.13} & { } & {}  \\ 
 & RI & 53.87 ± 0.13  & {91.29 ± 0.07} & {90.88 ± 0.06} & {93.57 ± 0.15} & { } & {}  \\ 
 & RS & 87.73 ± 0.15 & {92.56 ± 0.02} & {91.44 ± 0.11} & {89.02 ± 0.08} & { } & {}  \\  
  & RD & 73.59 ± 0.08 &{92.87 ± 0.05} & {92.01 ± 0.05} & {92.21 ± 0.16} & { } & {}  \\   
 & BT & 73.24 ± 0.11  & {86.05 ± 0.11} & {93.14 ± 0.04} & {90.25 ± 0.12} & { } & {}  \\  
 & MixCode & {88.13 ± 0.13}& {95.24 ± 0.12} & {94.05 ± 0.15} & {-} & {} & \\  
\multirow{-8}{*}{GraphCodeBERT} &  GenCode & \cellcolor[HTML]{DBE7FC}{89.78 ± 0.16} & \cellcolor[HTML]{DBE7FC}{95.67 ± 0.15} & \cellcolor[HTML]{DBE7FC}{94.89 ± 0.08} & \cellcolor[HTML]{DBE7FC}{93.89 ± 0.11} &  & \\    \cline{1-6}

 & No Aug  & {89.31 ± 0.15} & {92.89 ± 0.08} & {91.13 ± 0.12} & {91.93 ± 0.13} & {}  \\ 

 & SR & 68.94 ± 0.06 & {90.21 ± 0.12} & {88.23 ± 0.08} & {89.43 ± 0.08} & { } & {}  \\ 
 & RI & 58.34 ± 0.18 & {91.14 ± 0.14} & {91.13 ± 0.13} & {93.29 ± 0.17} & { } & {}  \\ 
 & RS & 89.98 ± 0.19 & {92.89 ± 0.13} & {90.87 ± 0.07} & {88.34 ± 0.07} & { } & {}  \\  
  & RD & 75.57 ± 0.13  & {93.98 ± 0.12} & {91.97 ± 0.11} & {92.35 ± 0.14} & { } & {}  \\   
 & BT & 75.64 ± 0.11 & {88.79 ± 0.08} & {94.06 ± 0.16} & {90.87 ± 0.05} & { } & {}  \\  
 & MixCode & {88.45 ± 0.14}& {95.86 ± 0.14} & {93.89 ± 0.19} & {-} & {} & \\  
\multirow{-8}{*}{CodeT5} &  GenCode & \cellcolor[HTML]{DBE7FC}{91.38 ± 0.09} & \cellcolor[HTML]{DBE7FC}{95.87 ± 0.11} & \cellcolor[HTML]{DBE7FC}{95.06 ± 0.12} & \cellcolor[HTML]{DBE7FC}{94.03 ± 0.12} & {} & \\    \cline{1-6}

 & No Aug  & {92.53 ± 0.02} & {94.12 ± 0.05} & {92.71 ± 0.11} & {93.21 ± 0.04} & {}  \\ 

 & SR & 82.75 ± 0.03 & {92.31 ± 0.07} & {91.87 ± 0.03} & {91.45 ± 0.05} & { } & {}  \\ 
 & RI & 75.89 ± 0.05 & {93.19 ± 0.12} & {92.35 ± 0.05} & {93.88 ± 0.13} & { } & {}  \\ 
 & RS & 92.13 ± 0.06 & {94.11 ± 0.08} & {91.67 ± 0.14} & {90.11 ± 0.06} & { } & {}  \\  
  & RD & 90.15 ± 0.08 & {94.67 ± 0.09} & {93.03 ± 0.02} & {92.89 ± 0.09} & { } & {}  \\   
 & BT & 85.62 ± 0.04  & {90.12 ± 0.13} & {92.55 ± 0.09} & {91.12 ± 0.03} & { } & {}  \\  
 & MixCode & {92.62 ± 0.01}& {95.91 ± 0.01} & {92.58 ± 0.06} & {-} & {} & \\  
\multirow{-8}{*}{StarCoder2-7B} &  GenCode & \cellcolor[HTML]{DBE7FC}{92.91 ± 0.03} &  \cellcolor[HTML]{DBE7FC}{96.32 ± 0.12} &  \cellcolor[HTML]{DBE7FC}{93.03 ± 0.04} & \cellcolor[HTML]{DBE7FC}{94.21 ± 0.12} & {} & \\    \cline{1-6}

 & No Aug  & {92.61 ± 0.03} & {94.35 ± 0.11} & {92.82 ± 0.14}  & {93.45 ± 0.06} & {}  \\ 

 & SR & 85.67 ± 0.04 & {92.67 ± 0.13} & {91.93 ± 0.08} & {91.55 ± 0.08} & { } & {}  \\ 
 & RI & 78.94 ± 0.06 & {93.01 ± 0.21} & {92.65 ± 0.06} & {93.75 ± 0.04} & { } & {}  \\ 
 & RS & 91.59 ± 0.07 & {94.31 ± 0.09} & {92.04 ± 0.02} & {90.32 ± 0.09} & { } & {}  \\  
 & RD & 90.65 ± 0.11 & {95.01 ± 0.04} & {92.65 ± 0.07} & {93.14 ± 0.12} & { } & {}  \\   
 & BT & 83.65 ± 0.02 & {91.23 ± 0.11} & {92.86 ± 0.13} & {91.56 ± 0.05} & { } & {}  \\  
 & MixCode & {92.89 ± 0.09 }& {94.89 ± 0.02} & {92.67 ± 0.03} & {-} & {} & \\  
\multirow{-8}{*}{Qwen2.5-Coder-7B} &  GenCode & \cellcolor[HTML]{DBE7FC}{93.12 ± 0.02} & \cellcolor[HTML]{DBE7FC}{96.17 ± 0.15} & \cellcolor[HTML]{DBE7FC}{92.93 ± 0.09} & \cellcolor[HTML]{DBE7FC}{94.38 ± 0.14} & {} & \\    \cline{1-6}
\end{tabular}
}

\vspace{-3mm}
\end{table*}

\textbf{Maximum probability analysis.} To explore why GenCode produces a more robust code model, we check the \emph{confidence} of trained models using different code augmentation methods on the original test data. Normally, if the DNN model is more confident in the test data, it is more difficult to attack this model using this data~\cite {pmlr-v80-wu18e}. To do so, for each dataset and code model, we collect the maximum probability of the corrected classified test data produced by the code model first. Then, we calculate the mean values of these maximum probability scores. The results of the mean values of maximum probabilities are shown in Table~\ref{tab:probability}. We can see that code models trained by GenCode are more confident in the test data than models trained using other code augmentation methods. As a result, GenCode trained code models are harder to attack.
\begin{table}[h]
\caption{Mean values of
maximum probabilities. The best results are highlighted in blue.  Code understanding tasks include \textbf{Authorship attribution} (GCJ),  \textbf{Bug detection} (Refactory), \textbf{Problem classification} (Java250), and \textbf{Clone detection} (BigCloneBench). }
\label{tab:probability}
\centering
\resizebox{0.95\columnwidth}{!}{
\begin{tabular}{clcccccc}
 \cline{1-6}
\textbf{Model} & \textbf{DA method} & \textbf{GCJ}  & \textbf{Refactory} & \textbf{Java250} & \textbf{BigCloneBench} & \\ \cline{1-6} 
 & No Aug  & {0.0381945} & {0.9368535} & {0.9554731} & {0.9651272} & {} & \\ 
 &Refactor & {0.0391864} & {0.9643155} & {0.9620814} & {0.9654681} & {} & \\ 
 & SR & {0.0380562} & {0.9527686} & {0.9341456} & {0.9583812} & {} & \\ 
 & RI & {0.0371215} & {0.9189242} & {0.9327171} & {0.9704512} & {} & \\  
 & RS & {0.0392987}  & {0.9561591} & {0.9581767} & {0.9410937} & {} & \\  
  & RD & {0.0392987}  & {0.9723022} & {0.9634647} & {0.9631375} & {} & \\  
 & BT & {0.0379581} & {0.9566747} & {0.9417861} & {0.9667154} & {} & \\  
 & MixCode & {0.0390714}& {0.9687351} & {0.9685594} & {-} & {} & \\  
\multirow{-9}{*}{CodeBERT} &  GenCode & \cellcolor[HTML]{DBE7FC}{0.0413779} & \cellcolor[HTML]{DBE7FC}{0.9808122} & \cellcolor[HTML]{DBE7FC}{0.9714346} & \cellcolor[HTML]{DBE7FC}{0.9786421} & {} & \\    
\cline{1-6} 
\end{tabular}
}
\vspace{-3mm}
\end{table}

\textbf{Statistical analysis.} We also conduct statistical tests using \emph{Wilcoxon signed-rank test}. When comparing existing data augmentation methods, GenCode consistently exhibits a \emph{p-value} $<$ 0.05 across all pre-trained code models shown from the lower of Table~\ref{table:statistical_testing_acc_robust}. In conclusion, GenCode can produce more adversarial robust code understanding models than other methods. 

\begin{tcolorbox}
\textbf{Answer to RQ2}: Our statistical tests find GenCode can train more robust pre-trained code models than existing data augmentation methods. On average, code models trained using GenCode have 8.42\% and 4.90\% higher robustness than models trained without data augmentation and models trained using MixCode, respectively. However, for LLMs, the robustness improvement achieved by GenCode remains limited, with gains of less than 1\%.
\end{tcolorbox}

\subsection{RQ3: How does the influence score affect the effectiveness of GenCode?}
\label{sec:results_ablation}

Even though we have preliminarily shown that selecting the code samples with higher loss values is a valid way to train good models, it is still necessary to study to what extent the advantage can be brought from this selection strategy. Therefore, in this part, we conduct an ablation study to compare GenCode with its two variants, GenCode with data selection using minimum loss values - GenCode (Min)  and random selection - GenCode (Random), respectively.

\begin{table}[h]
\caption{Effectiveness of data augmentation methods w.r.t. test accuracy $\uparrow$ (average $\pm$ standard deviation, \%) on original test data. The best results are highlighted in blue.  Code understanding tasks include \textbf{Authorship attribution} (GCJ),  \textbf{Bug detection} (Refactory), \textbf{Problem classification} (Java250), and \textbf{Clone detection} (BigCloneBench).}
\label{tab:ACC_ablation}
\centering
\resizebox{.95\columnwidth}{!}{
\begin{tabular}{clcccccc}
\cline{1-6}
\textbf{Model} & \textbf{DA method} & \textbf{GCJ}  & \textbf{Refactory} & \textbf{Java250} & \textbf{BigCloneBench} & \\ \cline{1-6} 
 & Random  & {94.74 ± 0.21} & {98.82 ± 0.15} & {98.24 ± 0.11} & {98.27 ± 0.19} & {} & \\ 
 & Min & {93.34 ± 0.32} & {96.13 ± 0.21} & {96.43 ± 0.17} & {97.23 ± 0.14} & {} & \\ 
\multirow{-3}{*}{CodeBERT} &  Max & \cellcolor[HTML]{DBE7FC}{96.23 ± 0.18} & \cellcolor[HTML]{DBE7FC}{99.32 ± 0.24} & \cellcolor[HTML]{DBE7FC}{98.89 ± 0.09} & {\cellcolor[HTML]{DBE7FC}98.93 ± 0.15} & {} & \\    \cline{1-6} 

 & Random & 95.26 ± 0.11 & {98.93 ± 0.18} & {98.68 ± 0.22} & {98.67 ± 0.13} & {} & \\ 
 & Min & {93.53 ± 0.17} & {96.31 ± 0.24} & {96.72 ± 0.12} & {97.22 ± 0.21} & {} & \\  
\multirow{-3}{*}{GraphCodeBERT} & Max & \cellcolor[HTML]{DBE7FC}{97.93 ± 0.14} & \cellcolor[HTML]{DBE7FC}{99.45 ± 0.17} & \cellcolor[HTML]{DBE7FC}{99.16 ± 0.13} & \cellcolor[HTML]{DBE7FC}{99.23 ± 0.09} & {} & \\   \cline{1-6}

 & Random & 96.24 ± 0.21 & {98.99 ± 0.21} & {98.93 ± 0.16} & {98.83 ± 0.08} & {} & \\ 
 & Min  & {94.11  ±  0.31} & {98.29 ± 0.12} & {97.03 ± 0.23} & {97.01 ± 0.15} & {} & \\  
\multirow{-3}{*}{CodeT5} & Max & \cellcolor[HTML]{DBE7FC}{98.85 ± 0.16} & \cellcolor[HTML]{DBE7FC}{99.64 ± 0.37} & \cellcolor[HTML]{DBE7FC}{99.38 ± 0.11} & \cellcolor[HTML]{DBE7FC}{99.34 ± 0.12} &  & \\ \cline{1-6}

 & Random & 96.81 ± 0.06  & {99.11 ± 0.02} & {99.23 ± 0.04} & {99.31 ± 0.06} & {} & \\ 
 & Min  & {95.42 ± 0.03} & {98.68 ± 0.04} & {98.47 ± 0.03} & {98.13 ± 0.03} & {} & \\  
\multirow{-3}{*}{StarCoder2-7B} & Max & \cellcolor[HTML]{DBE7FC}{98.93 ± 0.04 } & \cellcolor[HTML]{DBE7FC}{99.65 ± 0.02} & \cellcolor[HTML]{DBE7FC}{99.41 ± 0.05} & \cellcolor[HTML]{DBE7FC}{99.36 ± 0.09} &  & \\ \cline{1-6}

 & Random & 97.12 ± 0.05  & {99.12 ± 0.03} & {99.23 ± 0.03} & {99.32 ± 0.04} & {} & \\ 
 & Min  & {95.86 ± 0.08} & {98.71 ± 0.01} & {98.51 ± 0.05} & {98.19 ± 0.07} & {} & \\  
\multirow{-3}{*}{Qwen2.5-Coder-7B} & Max & \cellcolor[HTML]{DBE7FC}{98.96 ± 0.04} & \cellcolor[HTML]{DBE7FC}{99.65 ± 0.01} & \cellcolor[HTML]{DBE7FC}{99.42 ± 0.04} & \cellcolor[HTML]{DBE7FC}{99.36 ± 0.05} &  & \\ \cline{1-6}

\end{tabular}
}
\vspace{-3mm}
\end{table}

Table~\ref{tab:ACC_ablation} and Table~\ref{tab:Robustness_ablation} summarize the accuracy and robustness of trained code models using different selection strategies. We can see GenCode consistently outperforms the two variants under all testing criteria. Selecting code samples with minimum loss values is the worst way for GenCode. Interestingly, we found that GenCode~(Random) performs better than all baseline methods (i.e., \emph{No Aug}, Refactor, and MixCode) in terms of both accuracy and robustness improvements based on the results.

\begin{table}[h]
\caption{Effectiveness of data augmentation methods w.r.t. ASR $\downarrow$ (\%) on test data. The best results are highlighted in blue.  Code understanding tasks include \textbf{Authorship attribution} (GCJ),  \textbf{Bug detection} (Refactory), \textbf{Problem classification} (Java250), and \textbf{Clone detection} (BigCloneBench).}
\centering
\resizebox{0.95\columnwidth}{!}{
\begin{tabular}{clcccccccccccc}
\hline
 & \multicolumn{1}{c}{} & \multicolumn{2}{c}{\textbf{GCJ}} & \multicolumn{2}{c}{\textbf{Refactory}} & \multicolumn{2}{c}{\textbf{Java250}} & \multicolumn{2}{c}{\textbf{BigCloneBench}} \\
\multirow{-2}{*}{\textbf{Model}} & \multicolumn{1}{l}{\multirow{-2}{*}{\textbf{DA method}}} & \textbf{MHM} & \textbf{ALERT} & \textbf{\textbf{MHM}} & \textbf{ALERT} & \textbf{MHM} & \textbf{ALERT} & \textbf{MHM} & \textbf{ALERT}  \\ \hline
 & Random & \multicolumn{1}{c}{{25.21}} & {50.32} & \multicolumn{1}{c}{{27.65}} & {36.65} & \multicolumn{1}{c}{32.52} & {44.87} & \multicolumn{1}{c}{8.01} & {18.12} \\ 
 & Min & \multicolumn{1}{c}{{35.66}} & {66.67} & \multicolumn{1}{c}{{65.51}} & {69.76} & \multicolumn{1}{c}{46.76} & {53.86} & \multicolumn{1}{c}{20.65} & {41.32} \\ 
\multirow{-3}{*}{CodeBERT} & Max & \multicolumn{1}{c}{\cellcolor[HTML]{DBE7FC}{22.33}} & \cellcolor[HTML]{DBE7FC}{42.62} & \multicolumn{1}{c}{\cellcolor[HTML]{DBE7FC}{22.75}} & \cellcolor[HTML]{DBE7FC}{29.25} & \cellcolor[HTML]{DBE7FC}{29.89} & \cellcolor[HTML]{DBE7FC}{40.65} & \cellcolor[HTML]{DBE7FC}{7.12} &  \cellcolor[HTML]{DBE7FC}{15.83} \\ \hline

 & Random & \multicolumn{1}{c}{{29.97}} & {50.63} & \multicolumn{1}{c}{17.56} & {24.29} & \multicolumn{1}{c}{20.65} & {38.86} & \multicolumn{1}{c}{2.04} & {5.97} \\ 
 & Min & \multicolumn{1}{c}{{40.55}} & {68.23} & \multicolumn{1}{c}{{38.73}} & {46.98} & \multicolumn{1}{c}{26.87} & {46.67} & \multicolumn{1}{c}{11.45} & {21.23} \\ 
\multirow{-3}{*}{GraphCodeBERT} & Max & \multicolumn{1}{c}{\cellcolor[HTML]{DBE7FC}{26.43}} & \cellcolor[HTML]{DBE7FC}{39.87} & \multicolumn{1}{c}{\cellcolor[HTML]{DBE7FC}{15.78}} & \cellcolor[HTML]{DBE7FC}{20.92} & \cellcolor[HTML]{DBE7FC}{18.76} & \cellcolor[HTML]{DBE7FC}{33.81} & \cellcolor[HTML]{DBE7FC}{1.86} & \cellcolor[HTML]{DBE7FC}{4.62} \\ \hline

 & Random & \multicolumn{1}{c}{{25.03}} & {47.61} & \multicolumn{1}{c}{18.46} & {23.67} & \multicolumn{1}{c}{25.11} & {42.34} & \multicolumn{1}{c}{8.67} & {17.42} \\ 
 & Min & \multicolumn{1}{c}{{38.65}} & {59.89} & \multicolumn{1}{c}{35.67} & {48.59} & \multicolumn{1}{c}{32.61} & {49.27} & \multicolumn{1}{c}{18.36} & {25.54} \\ 
\multirow{-3}{*}{CodeT5} & Max & \multicolumn{1}{c}{\cellcolor[HTML]{DBE7FC}{23.57}} & \cellcolor[HTML]{DBE7FC}{38.76} & \multicolumn{1}{c}{\cellcolor[HTML]{DBE7FC}{12.13}} & \cellcolor[HTML]{DBE7FC}{18.81} & \cellcolor[HTML]{DBE7FC}{21.63} & \cellcolor[HTML]{DBE7FC}{39.76} & \cellcolor[HTML]{DBE7FC}{7.39} & \cellcolor[HTML]{DBE7FC}{12.15} \\ \hline

\end{tabular}}
\label{tab:Robustness_ablation}
\vspace{-3mm}
\end{table}

This phenomenon indicates that simply combining all syntax-preserving methods including code refactoring methods and syntax-breaking including text-oriented methods, is already a good code ugmentation method.

\begin{tcolorbox}
\textbf{Answer to RQ3}: Selecting code samples with maximum loss values is the best strategy among our considered variants, i.e., with minimum loss values and random selection. Besides, GenCode with random selection can outperform existing code augmentation methods.
\end{tcolorbox}

\subsection{RQ4: How does the data selection method affect the effectiveness of GenCode?}
\label{sec:results_DA_selections}

Table~\ref{tab:ACC_ablation_selection} presents the results of accuracy. Overall, LIME demonstrates limited effectiveness and generally underperforms compared to the gradient-based method (i.e., ~\emph{BADGE}). \emph{Influence Function} achieves better performance in scenarios involving smaller datasets, such as CodeT5 on GCJ and GraphCodeBERT on Refactory. Additionally, \emph{BADGE} exhibits superior performance on LLMs, particularly in tasks such as Qwen2.5-Coder-7B on Refactory and BigCloneBench. Although GenCode outperforms other methods in 12 out of 20 cases, these results suggest that exploring combinations of existing data selection strategies could be a promising direction for future research.
\begin{table}[h]
\caption{Effectiveness of data augmentation methods w.r.t. test accuracy $\uparrow$ (average $\pm$ standard deviation, \%) on original test data. The best results are highlighted in blue.  Code understanding tasks include \textbf{Authorship attribution} (GCJ),  \textbf{Bug detection} (Refactory), \textbf{Problem classification} (Java250), and \textbf{Clone detection} (BigCloneBench).}
\label{tab:ACC_ablation_selection}
\centering
\resizebox{.95\columnwidth}{!}{
\begin{tabular}{clcccccc}
\cline{1-6}
\textbf{Model} & \textbf{Data Selection} & \textbf{GCJ}  & \textbf{Refactory} & \textbf{Java250} & \textbf{BigCloneBench} & \\ \cline{1-6} 
 & LIME  & {93.67 ± 0.21} & {98.81 ± 0.13} & {97.25 ± 0.19} & {97.45 ± 0.14} & {} & \\ 
 & Influence Function  & {95.65 ± 0.19} & \cellcolor[HTML]{DBE7FC}{99.38 ± 0.09} & {97.13 ± 0.24} & {97.58 ± 0.19} & {} & \\ 
  & BADGE  & {96.15 ± 0.13} & {99.34 ± 0.22} & {98.21 ± 0.17} & \cellcolor[HTML]{DBE7FC}{98.95 ± 0.12} & {} & \\  
\multirow{-4}{*}{CodeBERT} &  GenCode & \cellcolor[HTML]{DBE7FC}{96.23 ± 0.18} & {99.32 ± 0.24} & \cellcolor[HTML]{DBE7FC}{98.89 ± 0.09} & {98.93 ± 0.15} & {} & \\    \cline{1-6} 

 & LIME  & {94.98 ± 0.07} & {98.12 ± 0.17 } & {98.01 ± 0.09} & {97.83  ± 0.09} & {} & \\ 
 & Influence Function  & \cellcolor[HTML]{DBE7FC}{97.98 ± 0.11} & \cellcolor[HTML]{DBE7FC}{99.46 ± 0.02} & {98.79  ± 0.08} & {97.79 ± 0.13} & {} & \\ 
  & BADGE  & {96.88 ± 0.07} & {98.65 ± 0.06} & {99.03 ± 0.11} & {98.41 ± 0.04} & {} & \\  
\multirow{-4}{*}{GraphCodeBERT} & GenCode & {97.93 ± 0.14} & {99.45 ± 0.17} & \cellcolor[HTML]{DBE7FC}{99.16 ± 0.13} & \cellcolor[HTML]{DBE7FC}{99.23 ± 0.09} & {} & \\   \cline{1-6}

 & LIME  & {96.36 ± 0.11} & {98.45 ± 0.06} & {99.03 ± 0.04} & {98.21 ± 0.13} & {} & \\ 
 & Influence Function  & \cellcolor[HTML]{DBE7FC}{98.97 ± 0.09} & {99.23 ± 0.13} & {99.29 ± 0.13} & {99.31 ± 0.12} & {} & \\ 
  & BADGE  & {98.86 ± 0.18} & {99.54 ± 0.16} & {99.36 ± 0.03} & {99.33 ± 0.18} & {} & \\  
\multirow{-4}{*}{CodeT5} & GenCode & {98.85 ± 0.16} & \cellcolor[HTML]{DBE7FC}{99.64 ± 0.37} & \cellcolor[HTML]{DBE7FC}{99.38 ± 0.11} & \cellcolor[HTML]{DBE7FC}{99.34 ± 0.12} &  & \\ \cline{1-6}

 & LIME  & {96.96 ± 0.08} & {99.13 ± 0.04} & {99.25 ± 0.09} & {99.32 ± 0.04} & {} & \\ 
 & Influence Function  & {97.35 ± 0.05} & {99.41 ± 0.06} & {99.31 ± 0.05} & {99.34 ± 0.03} & {} & \\ 
  & BADGE  & {98.96 ± 0.14} & \cellcolor[HTML]{DBE7FC}{99.65 ± 0.04} & {99.39 ± 0.12} & {99.35 ± 0.06} & {} & \\    
\multirow{-4}{*}{StarCoder2-7B} & GenCode & \cellcolor[HTML]{DBE7FC}{98.93 ± 0.04} & \cellcolor[HTML]{DBE7FC}{99.65 ± 0.02} & \cellcolor[HTML]{DBE7FC}{99.41 ± 0.05} & \cellcolor[HTML]{DBE7FC}{99.36 ± 0.09} &  & \\ \cline{1-6}

  & LIME  & {97.46 ± 0.05} & {99.25 ± 0.12 } & {99.04 ± 0.02} & {99.32 ± 0.13} & {} & \\ 
 & Influence Function  & {98.13 ± 0.07} & {99.52 ± 0.05} & {99.12 ± 0.07} & {99.35 ± 0.02} & {} & \\ 
  & BADGE  & \cellcolor[HTML]{DBE7FC}{98.98 ± 0.06} & {99.49 ± 0.03} & {99.38 ± 0.02} & \cellcolor[HTML]{DBE7FC}{99.37 ± 0.03} & {} & \\  
\multirow{-4}{*}{Qwen2.5-Coder-7B} & GenCode  & {98.96 ± 0.04} & \cellcolor[HTML]{DBE7FC}{99.65 ± 0.01} & \cellcolor[HTML]{DBE7FC}{99.42 ± 0.04} & {99.36 ± 0.05} &  & \\ \cline{1-6}

\end{tabular}
}
\vspace{-3mm}
\end{table}

\begin{table}[h]
\caption{Effectiveness of data augmentation methods w.r.t. test accuracy $\uparrow$ (average $\pm$ standard deviation, \%) on transformed test data. The best results are highlighted in blue.  Code understanding tasks include \textbf{Authorship attribution} (GCJ),  \textbf{Bug detection} (Refactory), \textbf{Problem classification} (Java250), and \textbf{Clone detection} (BigCloneBench).}
\label{tab:Robust_ablation_selection}
\centering
\resizebox{.95\columnwidth}{!}{
\begin{tabular}{clcccccc}
\cline{1-6}
\textbf{Model} & \textbf{Data Selection} & \textbf{GCJ}  & \textbf{Refactory} & \textbf{Java250} & \textbf{BigCloneBench} & \\ \cline{1-6} 
 & LIME  & {89.13 ± 0.07} & {95.41 ± 0.11} & {93.21 ± 0.14} & {91.68 ± 0.12} & {} & \\ 
 & Influence Function  & {90.34 ± 0.05} & {95.51 ± 0.07} & {93.35 ± 0.12} & {92.17 ± 0.15} & {} & \\ 
  & BADGE  & \cellcolor[HTML]{DBE7FC}{91.31 ± 0.09} & {95.61 ± 0.05} & {93.61 ± 0.03} & \cellcolor[HTML]{DBE7FC}{92.64 ± 0.11} & {} & \\  
\multirow{-4}{*}{CodeBERT} &  GenCode & {91.23 ± 0.14} & \cellcolor[HTML]{DBE7FC}{95.63 ± 0.14} & \cellcolor[HTML]{DBE7FC}{93.62 ± 0.09} & {92.56 ± 0.15} & {} & \\    \cline{1-6} 

 & LIME  & {88.24 ± 0.06} & {95.35 ± 0.07} & {94.24 ± 0.02} & {93.62 ± 0.08} & {} & \\ 
 & Influence Function  & {88.97 ± 0.08} & {95.62 ± 0.03} & {94.32 ± 0.07} & {93.58 ± 0.12} & {} & \\ 
  & BADGE  & \cellcolor[HTML]{DBE7FC}{90.01 ± 0.04} & \cellcolor[HTML]{DBE7FC}{95.88 ± 0.05} & {94.87 ± 0.11} & \cellcolor[HTML]{DBE7FC}{93.96 ± 0.07} & {} & \\  
\multirow{-4}{*}{GraphCodeBERT} &  GenCode & {89.78 ± 0.16} & {95.67 ± 0.15} & \cellcolor[HTML]{DBE7FC}{94.89 ± 0.08} & {93.89 ± 0.11} &  & \\   \cline{1-6}

 & LIME  & {90.45 ± 0.13} & {95.85 ± 0.13} & {94.32 ± 0.04} & {93.43 ± 0.05} & {} & \\ 
 & Influence Function  & {91.04 ± 0.16} & {95.94 ± 0.11} & {94.68 ± 0.09} & {93.65 ± 0.09} & {} & \\ 
  & BADGE  & {91.35 ± 0.14} & \cellcolor[HTML]{DBE7FC}{96.14 ± 0.15} & \cellcolor[HTML]{DBE7FC}{95.14 ± 0.09} & {93.95 ± 0.14} & {} & \\  
\multirow{-4}{*}{CodeT5} & GenCode & \cellcolor[HTML]{DBE7FC}{91.38 ± 0.09} & {95.87 ± 0.11} & {95.06 ± 0.12} & \cellcolor[HTML]{DBE7FC}{94.03 ± 0.12} & {} & \\ \cline{1-6}

 & LIME  & {92.72 ± 0.16} & {94.56 ± 0.14} & {92.67 ± 0.12} & {93.91 ± 0.09} & {} & \\ 
 & Influence Function  & {92.89 ± 0.03} & {95.13 ± 0.19} & {92.86 ± 0.11} & {94.12 ± 0.15} & {} & \\ 
  & BADGE  & \cellcolor[HTML]{DBE7FC}{92.93 ± 0.04} & {96.24 ± 0.17} & \cellcolor[HTML]{DBE7FC}{93.12 ± 0.13} & {94.19 ± 0.19} & {} & \\    
\multirow{-4}{*}{StarCoder2-7B} & GenCode & {92.91 ± 0.03} &  \cellcolor[HTML]{DBE7FC}{96.32 ± 0.12} &  {93.03 ± 0.04} & \cellcolor[HTML]{DBE7FC}{94.21 ± 0.12} & {} & \\ \cline{1-6}

  & LIME  & {92.94 ± 0.07} & {95.13 ± 0.04} & {92.68 ± 0.12} & {93.87 ± 0.05} & {} & \\ 
 & Influence Function  & {92.88 ± 0.19} & {95.78 ± 0.09} & {92.76 ± 0.08} & {94.19 ± 0.06} & {} & \\ 
  & BADGE  & {93.09 ± 0.05} &  \cellcolor[HTML]{DBE7FC}{96.19 ± 0.02} & \cellcolor[HTML]{DBE7FC}{92.97 ± 0.11} & {94.32 ± 0.05} & {} & \\  
\multirow{-4}{*}{Qwen2.5-Coder-7B} &  GenCode & \cellcolor[HTML]{DBE7FC}{93.12 ± 0.02} & {96.17 ± 0.15} & {92.93 ± 0.09} & \cellcolor[HTML]{DBE7FC}{94.38 ± 0.14} & {} &  \\ \cline{1-6}

\end{tabular}
}
\vspace{-3mm}
\end{table}

An interesting observation from the natural robustness results in Table~\ref{tab:Robust_ablation_selection} is that, in most cases, the gradient-based method achieves better robustness improvements than GenCode. Specifically, \emph{BADGE} outperforms GenCode by integrating gradient-based uncertainty with diversity via K-means clustering, thereby facilitating the selection of informative and representative samples~\cite{ash2019deep}.

\begin{tcolorbox}
\textbf{Answer to RQ4}: GenCode achieves better improvements in accuracy compared to other data selection methods. However, the gradient-based method demonstrates superior performance in enhancing robustness. Investigating the trade-offs between GenCode and gradient-based methods presents an interesting direction for future research.
\end{tcolorbox}

\section{Discussion}

\subsection{Computational cost of augmented data}

\begin{figure*}[!tb]
\centering
\subfigure[Authorship attribution]{
\begin{minipage}[t]{0.5\linewidth}
\includegraphics[width=1.0\linewidth]{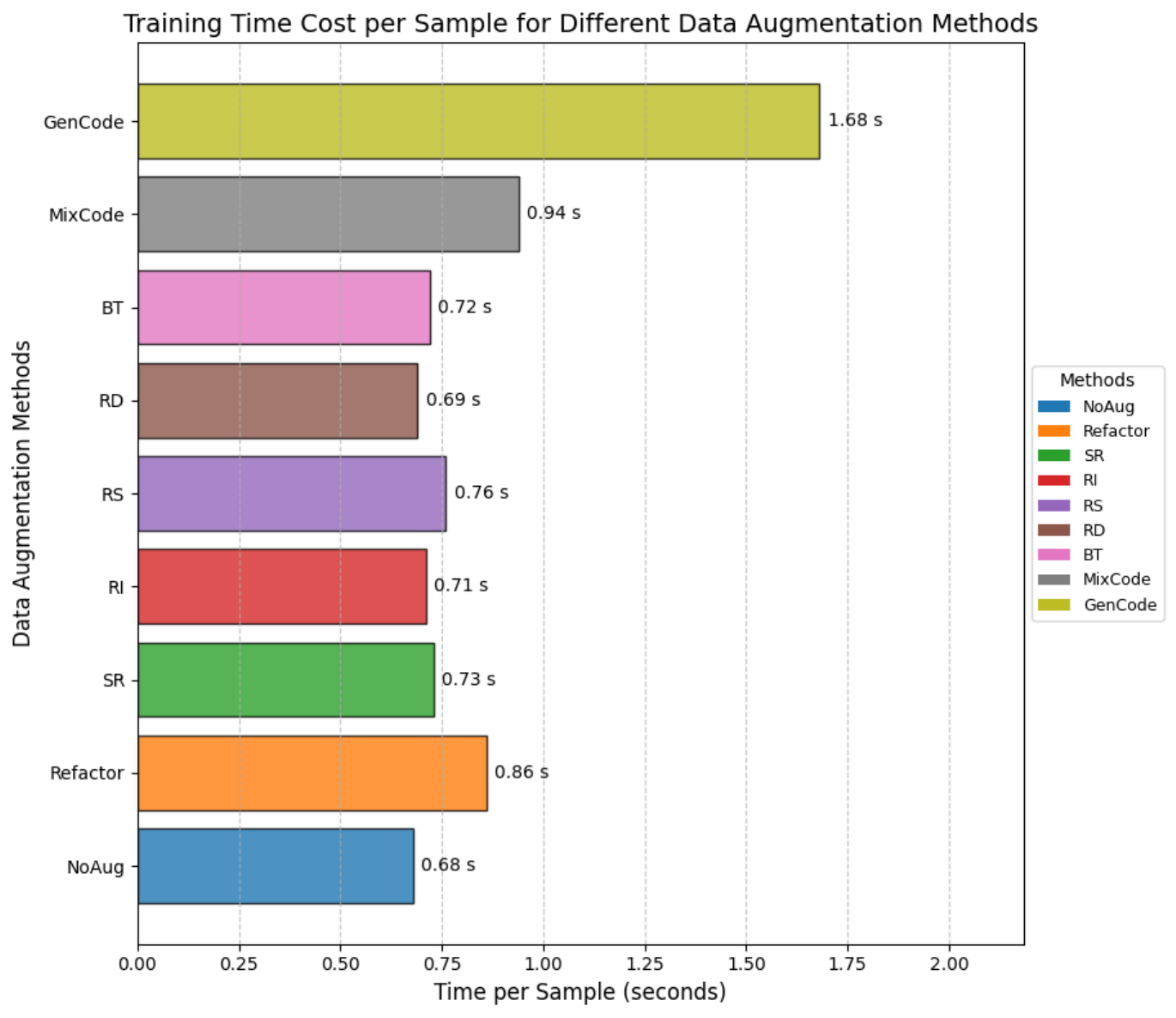}
\end{minipage}%
}%
\subfigure[Bug detection]{
\begin{minipage}[t]{0.5\linewidth}
\includegraphics[width=1.0\linewidth]{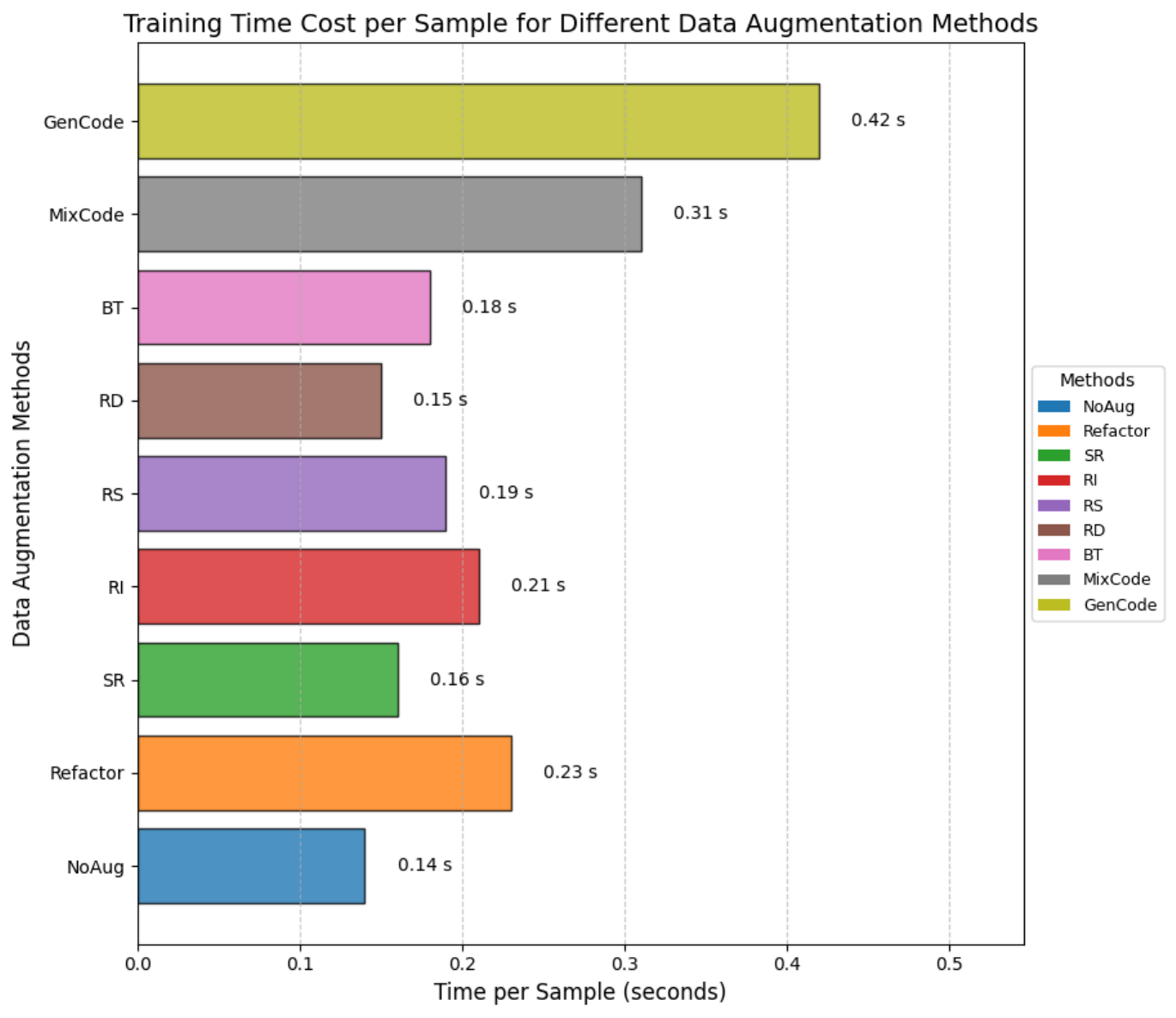}
\end{minipage}%
}%

\subfigure[Problem classification]{
\begin{minipage}[t]{0.5\linewidth}
\centering
\includegraphics[width=1.0\linewidth]{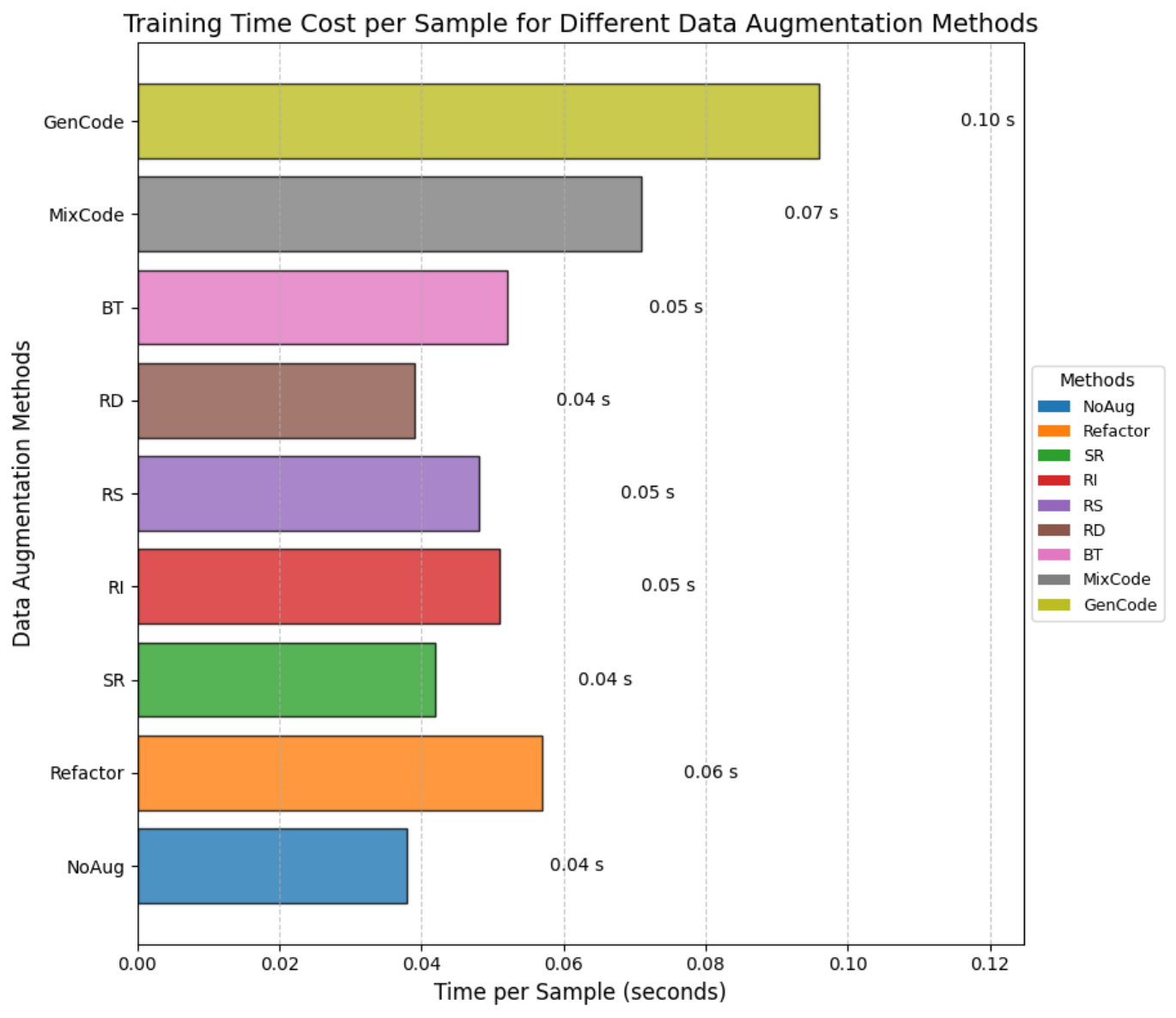}
\end{minipage}%
}%
\subfigure[Clone detection]{
\begin{minipage}[t]{0.5\linewidth}
\centering
\includegraphics[width=1.0\linewidth]{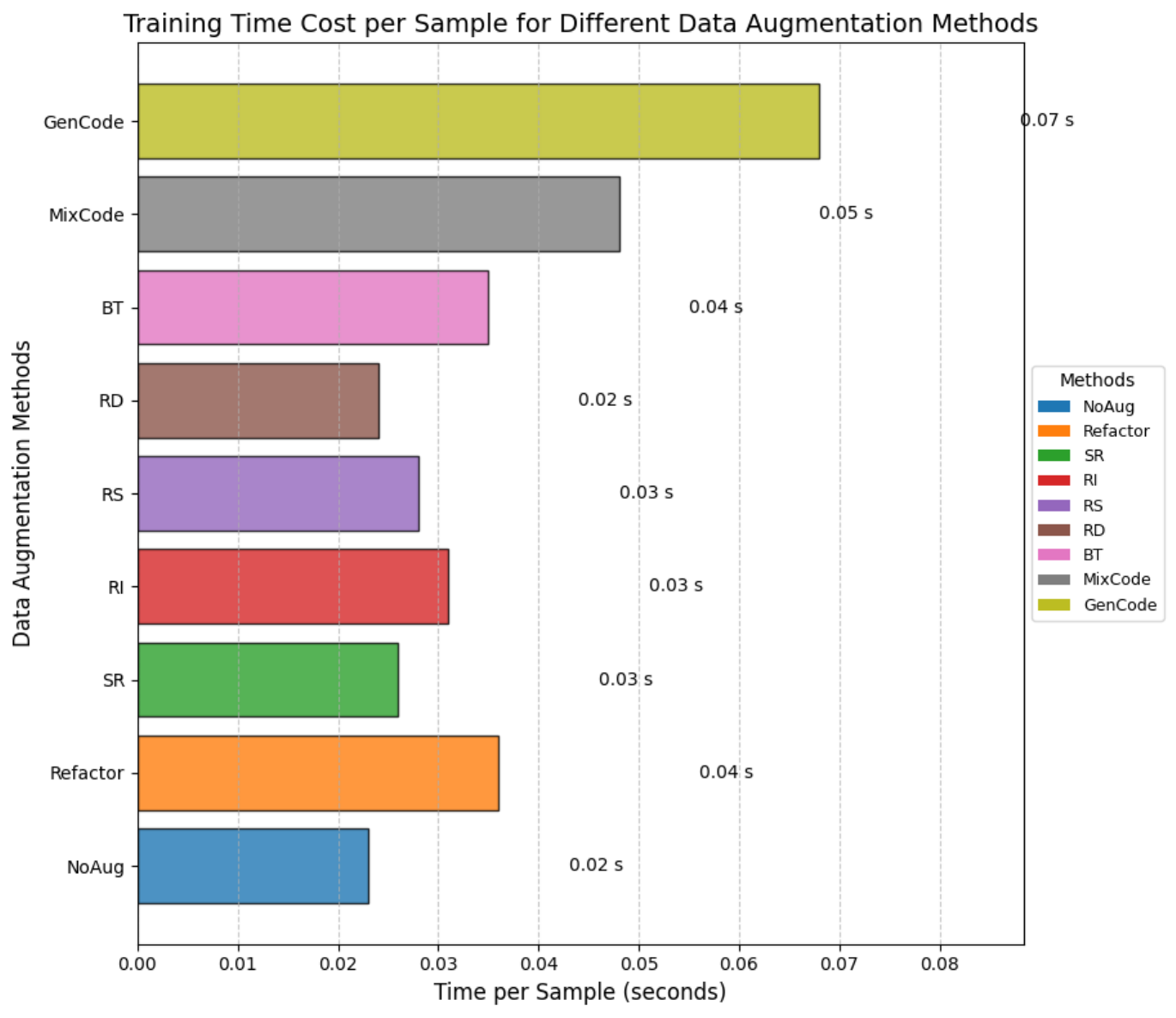}
\end{minipage}%
}%
\caption{Computational cost of CodeBERT using different code augmentation methods in each task.}
\label{fig:cost}
\end{figure*}

We conduct a case study to evaluate the computational cost of model training with different data augmentation methods. The results are visualized in Fig.~\ref{fig:cost}. We observe that GenCode incurs the highest computational cost compared to other data augmentation methods including code refactoring, text-oriented, and MixCode. Interestingly, text-oriented data augmentation methods consistently exhibit lower training-time cost than code refactoring methods. In particular, Random Deletion (RD) achieves the best trade-off between performance and efficiency.

Although Back Translation (BT) performs relatively well across some tasks (e.g., authorship attribution and bug detection), it requires significant preprocessing time to generate the augmented data before model training, making it less practical. Therefore, we do not recommend BT as a preferred augmentation strategy in time-sensitive scenarios.

In conclusion, rule-based augmentation methods (e.g., code refactoring or text-oriented data augmentation methods) typically require less computation during model training, but may incur the preprocessing overhead. In contrast, sample-based methods such as MixCode do not require data preparation in advance and can be flexibly applied during training. While GenCode consistently achieves the best accuracy and robustness, it comes with the highest computational cost. Exploring the trade-off between performance and efficiency is an important direction for future work.

\subsection{On the Data Distribution of Augmented Data}
One common concern in data augmentation is the risk of introducing samples that deviate from the original data distribution, potentially harming model generalization. While augmented data may not strictly follow the original distribution, prior studies (e.g., Mixup~\cite{zhang2017mixup}, CutMix~\cite{yun2019cutmix}) have shown that moderate, controlled deviations can actually enhance generalization. To mitigate distributional shift, GenCode follows a principled two-stage pipeline: (1) controlled data generation, and (2) loss-based importance-guided selection. In the first stage, GenCode applies a combination of semantic-preserving code refactorings (e.g., variable renaming, field enhancement) and lightweight text-oriented data augmentation methods(e.g., random swap, synonym replacement). These transformations are designed to preserve or minimally alter program semantics, ensuring that the generated data remains close to the original distribution.

In the second stage, GenCode evaluates each augmented sample based on its training loss under the current model. Only samples with the highest loss, those that are potentially more informative, are selected for training, while others are discarded. This loss-based filtering mechanism promotes alignment with the model’s effective learning region and helps avoid training on overly noisy or unrepresentative data.

\subsection{Limitations}
Even though the accuracy of our studied code models already achieved relatively high accuracy in these tasks, i.e., most of them have more than 90\% accuracy, their robustness is still low. Regarding the large language models (LLMs), we have checked the literature and found that the clone detection task is still challenging for LLMs, e.g., ChatGPT has only 0.784 Precision for Java-Java pair detection~\cite{khajezade2024investigating}. For other program understanding tasks that have a large number of classes, e.g., the Java250 dataset used in our experiments has 250 classes. It is also hard to use LLMs such as ChatGPT to predict the label since we need to add many label guides in the prompt.

\subsection{Threats to Validity}
The internal threat to validity mainly comes from the implementation of standard model training and data augmentation methods. The code that is used in our experiment includes 1) the authorship attribution task (Dataset: GCJ) taken from the existing work~\cite{yang2022natural}, 2) the bug detection task (Dataset: Refactory) adopted from~\cite{hu2019re}, 3) the program classification task (Dataset: Java250) sourced from Project CodeNeet~\cite{puri2021codenet}, 4) the clone detection task (Dataset: BigCloneBench) chosen from the work~~\cite{svajlenko2014towards}. The implementation of \emph{SR}, \emph{RI}, \emph{RS}, \emph{RD}, and \emph{BT} source from the newest release~\cite{marivate2020improving}, again we adapt them to our experiment. The implementation of \emph{MixCode} and \emph{Refactor} come from its original release~\cite{dong-mixcode}. 

The external threats to validity in our study are associated with the code understanding tasks, datasets, pre-trained code models, and data augmentation methods we have selected. We have considered four distinct code understanding tasks: authorship attribution, bug detection, problem classification, and clone detection, utilizing a total of four datasets for these tasks. Our study encompasses two widely used programming languages in the software community, namely Java and Python. Furthermore, we employ three types of pre-trained code models, namely \emph{CodeBERT}, \emph{GraphCodeBERT}, and \emph{CodeT5}, as well as two recently released code-specific large language models, StarCoder2 and Qwen2.5-Coder, in our experiments. For data augmentation methods from NLP, we have selected established techniques based on their widespread citation in academic papers and their recognition in authoritative journals and conferences.

The construct threats to validity in our study primarily come from various factors, including parameter settings, randomness, and the choice of evaluation measures. For MixCode, the primary parameter is lambda, which governs the weight of mixing two feature vectors. We have adhered to the original recommendation for this parameter. Similarly, for data augmentation methods from NLP, we have used the parameters as per their original release. To mitigate the influence of randomness, we have conducted each experiment five times and reported the results as an average along with the standard deviation. In terms of evaluation metrics, our study encompasses both accuracy and robustness. While accuracy assesses the basic performance of pre-trained code models, robustness is employed to evaluate their generalization capabilities.

\section{Related Work}
\label{sec:related}
Data augmentation has made remarkable success in the field of machine learning, e.g., natural language processing~\cite {feng2021survey} and code model~\cite {zhuo2023data}. Indeed, data augmentation has proven to be a valuable technique for improving model performance, specifically increasing trained model accuracy and enhancing dataset diversity~\cite{rebuffi2021data}. Inspired by that code can be represented as text data~\cite{allamanis2018survey}, we review related works from three perspectives, deep learning for code understanding, data augmentation for code models, and data augmentation for \emph{NLP}.

\subsection{Deep Learning for Code Understanding}

Code understanding is important for software developers during the development and maintenance phases. Benefiting from the potential of using deep learning models to understand natural language texts, researchers from the SE borrow the same idea to build deep learning models for code understanding.  Code2vec~\cite{alon2019code2vec} is the very first work that introduced the use of DNN models to learn vector information extracted by code snippets. Code2vec achieved promising results in understanding code on the function name prediction task. Later on, CodeBERT~\cite{feng2020codebert}, was proposed for more code-understanding tasks. CodeBERT follows the pre-training and fine-tuning paradigm. That means it uses multiple programming languages to pre-train the model to learn generation code information. Then, people can fine-tune the pre-trained CodeBERT on a specific downstream task for real usage. CodeBERT achieved impressive results on many code understanding such as comments generation. After that, multiple pre-trained code models have been proposed with ever better performance, such as GraphCodeBERT~\cite{guo2020graphcodebert}, GraphCode2Vec~\cite{ma2022graphcode2vec}, and CodeT5~\cite{wang-etal-2021-codet5}. More recently, large language models such as ChatGPT~\footnote{\url{https://chat.openai.com/}} have been demonstrated to be the best choice to solve code understanding tasks. 

Different from the above works that focused on designing new model architectures to understand code. We target a more general problem -- how to prepare the training data to further boost the training of code models. Our approach is suitable for any code understanding model. 

\subsection{Data Augmentation for code models}
Inspired by the success of data augmentation, researchers have recently dedicated significant efforts to applying data augmentation techniques to large-scale code-related tasks to enhance the performance of code models, focusing on accuracy and robustness. Adversarial training~\cite{goodfellow2014explaining}, which involves generating a set of adversarial examples and incorporating them into the training data, has been explored as a data augmentation method in code learning. To enhance the performance of deep comment generation models, Zhang \emph{et al.}~\cite{zhang2020training} proposed a code augmentation method that utilizes the Metropolis-Hastings Modifier (MHM) algorithm to generate the adversarial training dataset. Similar to their work, Mi \emph{et al.}~\cite{mi2021effectiveness} generated additional data using Auxiliary Classifier Generative Adversarial Networks (GANs). Besides, as a method that is tailored specifically for code-related tasks, code refactoring which involves restructuring and optimizing code without changing its external behavior is becoming a mainstream code augmentation.  Yu \emph{et al.}~\cite{yu2022data} devised program transformation rules tailored for Java and conducted an extensive assessment of their effectiveness when employed as code augmentation techniques across three significant code-related tasks. Allamanis \emph{et al.}~\cite{allamanis2021self} applied four simple code rewrite rules to improve the efficiency of model training in the program repair task. Furthermore, taking inspiration from \emph{Mixup}, which linearly mixes the features of two distinct image datasets, Dong~\emph{et al.}~\cite{dong-mixcode} proposed \emph{Mixcode} that linearly interpolates the features derived from a pair of programs along with their corresponding labels, wherein One of the programs undergoes code refactoring transformation, while the other remains unchanged. Li~\emph{et al.}~\cite{li-etal-2022-exploring-representation} also introduced alternative interpolation methods for the code model, namely binary interpolation and linear extrapolation. 
 
Unlike previous work, our study is the first one that considers the importance of code data for enhancing the performance of code understanding tasks. Specifically, during each epoch, the most valuable data selected by the metric score are fed into the code model for enhancing the performance of training. 

\subsection{Data Augmentation for NLP}
Data augmentation methods that are used in NLP can be roughly divided into three categories, namely \emph{paraphrasing}, \emph{noising-based methods}, and \emph{sampling-based methods}~\cite{feng2021survey}. Xie \emph{et al.}~\cite{xie2020unsupervised} proposed the back-translation technique that involves bidirectionally applying a translation model to each sentence. More precisely, this method first translates the original text into another language and then reverts it back to the original one. Paraphrasing is widely employed in the field of NLP since it can convey identical information as the original text but in a different form.  Wei \emph{et al.}~\cite{wei-zou-2019-eda} applied a series of nosing-based methods named \emph{EDA} to augment the text data. Noising-based methods generally involve the subtle addition of noise to the original data while preserving their semantic information. For example, \emph{Random Swap} (one of the methods from \emph{EDA}) can include randomly selecting two words in a sentence and swapping their positions. Besides, different from paraphrasing or noising-based methods, sampling-based methods are task-specific and necessitate both data and their label formats. Also inspired by the \emph{Mixup}, Guo \emph{et al.}~\cite{guo2019augmenting} adapted it from its original use with image data to text data. They conducted an empirical study to investigate its effectiveness. Specifically, they generated new synthetic data by linearly mixing the embeddings instead of directly operating on the raw text data. Furthermore, to overcome the limitations of traditional methods such as \emph{Synonym Replacement}, which generally require significant prior knowledge and are prone to sub-optimal results, Ren \emph{et al.}~\cite{ren2021taa} proposed \emph{Text AutoAugment}. In detail, it considers a combination of various text-oriented data augmentation operations as an augmentation policy and employs an efficient Bayesian Optimization algorithm to automatically search for the best policy.

In contrast to the data augmentation methods used in NLP, our work focuses on the strategy that guides code model training. The strategy involves both existing data augmentation methods derived directly from code data and those adapted from text data. Moreover, our approach also allows for the integration of new program transformation methods that researchers may develop in the future.

\section{Conclusion}
\label{sec:conclusion}
We propose the first generation-and-selection-based framework, GenCode, to boost code understanding models. The basic assumption behind GenCode is that less confident data contributes more to the model training. Based on this assumption, GenCode utilizes atomic data augmentation operators to generate candidate training data first and selects data with maximum loss values to train the code model. Our evaluation demonstrates that GenCode can produce more accurate and robust models than MixCode. Besides, we analyze how the influence score affects the effectiveness of GenCode.





\bibliographystyle{apalike}
\bibliography{reference}

\end{document}